\documentclass[iop]{emulateapj}

\usepackage{color}
\usepackage[backref,breaklinks,colorlinks,encap, linkcolor=blue,citecolor=blue,urlcolor=black]{hyperref}

\newcommand{\teff}{$T_{\text{eff}}$}

\shorttitle{Characterization of HAT-P-32A\MakeLowercase{b} and HAT-P-32B}
\shortauthors{Zhao et al.}

\begin{document}

\title{Characterization of the atmosphere of the hot Jupiter HAT-P-32A\MakeLowercase{b} and the M-dwarf companion HAT-P-32B}


\author{Ming Zhao\altaffilmark{1,2}, 
Joseph G. O'Rourke\altaffilmark{3} 
Jason T. Wright\altaffilmark{1,2},
Heather A. Knutson\altaffilmark{3},
Adam Burrows\altaffilmark{4},
Johnathan Fortney\altaffilmark{5},
Henry Ngo\altaffilmark{3},
Benjamin J. Fulton\altaffilmark{6},
Christoph Baranec\altaffilmark{6},
Reed Riddle\altaffilmark{7},
Nicholas M. Law\altaffilmark{8},
Philip S. Muirhead\altaffilmark{9},
Sasha Hinkley\altaffilmark{7},
Adam P. Showman\altaffilmark{10},
Jason Curtis \altaffilmark{1,2},
Rick Burruss\altaffilmark{11}
}


\altaffiltext{1}{Department of Astronomy \& Astrophysics, Pennsylvania State University, PA 16802, USA}
\altaffiltext{2}{Center for Exoplanets and Habitable Worlds}
\altaffiltext{3}{Division of Geological \& Planetary Sciences, California Institute of Technology, Pasadena, CA 91125, USA}
\altaffiltext{4}{Department of Astrophysical Sciences, Princeton University}
\altaffiltext{5}{Department of Astronomy and Astrophysics, University of California, Santa Cruz}
\altaffiltext{6}{Institute for Astronomy, University of Hawai`i at M\={a}noa, Hilo, HI 96720-2700, USA}
\altaffiltext{7}{Division of Physics, Mathematics, and Astronomy, California Institute of Technology, Pasadena, CA 91125, USA}
\altaffiltext{8}{Department of Physics and Astronomy, University of North Carolina at Chapel Hill, Chapel Hill, NC 27599-3255, USA}
\altaffiltext{9}{Department of Astronomy, Boston University, 725 Commonwealth Ave., Boston, MA  02215, USA}
\altaffiltext{10}{Department of Planetary Sciences and Lunar and Planetary Laboratory, The University of Arizona, Tucson, AZ 85721, USA}
\altaffiltext{11}{Jet Propulsion Laboratory, California Institute of Technology, CA 91109, USA}


\begin{abstract}
We report secondary eclipse photometry of the hot Jupiter HAT-P-32Ab, taken with Hale/WIRC in  $H$ and $K_S$ bands and with {\em Spitzer}/IRAC at 3.6 and 4.5 \micron. We carried out  adaptive optics imaging of the planet host star HAT-P-32A and its  companion HAT-P-32B in the near-IR and the visible. 
We  clearly resolve the two stars from each other and find a separation of 2\farcs923 $\pm$ 0\farcs004 and  a position angle 110\fdg64 $\pm$ 0\fdg12. 
We measure the flux ratios of the  binary in  $g'r'i'z'$  and  $H$ \& $K_S$ bands, and determine \teff~= 3565 $\pm$ 82 K for the companion star, corresponding to an M1.5 dwarf. 
We use PHOENIX stellar atmosphere models to 
correct the dilution of the secondary eclipse depths of the hot Jupiter due to the presence of the M1.5 companion.
We also improve the secondary eclipse photometry by accounting for the non-classical,  flux-dependent nonlinearity of the WIRC IR detector in the $H$ band.  We measure planet-to-star flux ratios of 0.090 $\pm$ 0.033\%, 0.178 $\pm$ 0.057\%, 0.364 $\pm$ 0.016\%, and 0.438 $\pm$ 0.020\% in the $H$, $K_S$, 3.6 and 4.5 \micron~bands, respectively. 
We compare these with planetary atmospheric models, and find they prefer an atmosphere with a temperature inversion and inefficient heat redistribution. 
However, we also find that the data are equally well-described by a blackbody model for the planet with $T_{\rm p} =2042\pm 50$ K. Finally, we measure a secondary eclipse timing offset of 0.3 $\pm$ 1.3 min from the predicted mid-eclipse time, which 
constrains $e$ = 0.0072$^{+0.0700}_{-0.0064}$  when combined with RV data and is more consistent with a circular orbit.
\end{abstract}





\section{Introduction}

Secondary eclipses (occultations) of transiting planets occur when the planet passes behind its host star.  Observations of these events in the infrared allow us to directly detect thermal emission from these planets, providing an unparalleled opportunity to study the chemistry and physics of exoplanetary atmospheres.
When measured at multiple wavelengths with high precision, the emission spectrum of a planet can be used to characterize planetary atmospheric temperature-pressure structure, chemistry, and heat recirculation \citep[e.g.,][]{Burrows et al.2006, Barman2008, Fortney et al.2008, Line et al.2013, Madhusudhan et al.2014}. 
To date, detections of thermal emission have been made for more than fifty planets, most of which were obtained using the {\em Spitzer Space Telescope}. 
{Since 2009, however, {\em Spitzer} has exhausted its cryogen and has been limited to observing in only the 3.6 and 4.5 \micron~bands.}
Ground-based observations have recently emerged as another important tool to measure secondary eclipses, providing highly complementary wavelength coverage to that of $Spitzer$ and even the {\em Hubble Space Telescope} \citep[e.g.,][]{Alonso et al.2009, Gillon2009, Gibson et al.2010, Croll et al.2010a, Croll et al.2010b, Croll et al.2011, Caceres et al.2011, Zhao et al.2012a, Zhao et al.2012b, Deming et al.2012, Bean et al.2013, Wang et al.2013, ORourke et al.2014, Shporer et al.2014, Chen et al.2014}. 
Because ground-based near-IR observations generally probe different layers of planetary atmospheres, they can provide important constraints and break degeneracies among differing temperature-pressure profiles and compositions \citep[e.g.,][]{Madhusudhan et al.2010, Madhusudhan et al.2011}. 

In addition to studying planetary atmospheres, the timing of the secondary eclipse relative to that of the primary transit also provides a tight constraint on $e \cos \omega$, where $e$ is the planet's orbital eccentricity and $\omega$ is the longitude of periastron.  {When combined with radial velocity observations, secondary eclipse timing data can reduce the uncertainty in the estimated eccentricity by a factor of $\sim$10 \citep[e.g.,][]{Lewis et al.2013, Knutson et al.2014}.}
 Precisely measured eccentricities allow for better estimates of planetary mass and radius \citep[e.g.,][]{Madhusudhan et al.2009b}, and can provide important information to the tidal circularization process. This may in turn shed light on the nature of the inflated radii observed in a subset of hot Jupiters \citep[e.g.,][]{Bodenheimer et al.2001, Miller et al.2009}.

A majority of the secondary eclipse observations obtained to date have focused on a class of short-period gas giant planets known as ``hot Jupiters".  This is due to their high temperatures and large radii, which result in particularly favorable planet-star flux ratios.
Previous studies of hot Jupiters have revealed that they are in fact a heterogeneous group. Some  planets seem to have a temperature inversion layer caused by unknown absorber in their upper atmospheres, while other planets seem to lack such an inversion layer \citep[e.g.,][]{Fortney et al.2008, Burrows et al.2008, Madhusudhan et al.2014, Line et al.2014}. \citet{Knutson et al.2010} found a correlation between stellar activity and the presence/absence of thermal inversions. They suggested that increased far UV flux from active stars might destroy the compounds responsible for the formation of the observed temperature inversions, preventing inversions from forming in planets orbiting chromospherically active stars. In addition, \citet{Madhusudhan2012} suggested that 
super-solar C/O ratios  (C/O $>$1) may explain the lack of inversions in some planets. 

Hot Jupiters experience strong irradiation from their host stars due to their close-in and tidally-locked orbits. The resulting heat on the dayside can be redistributed to the nightside by strong zonal winds (Showman et al. 2008). \citet{Cowan et al.2011} found that hot Jupiters with temperatures $\gtrsim$2400 K usually have very low global recirculation efficiency and large day-night temperature contrasts, while cooler planets have a wider variety of recirculation efficiencies.
\citet{Perez-Becker et al.2013} reproduced this observed trend  using a shallow-water model, and found that the transition between low and efficient heat redistribution depends on the timescale of gravity wave propagation, among other timescales. 

Among the large number of transiting exoplanets discovered to date, the hot Jupiter HAT-P-32Ab, discovered by \citet{Hartman et al.2011} with HATNet,  stands out as one of the three most inflated planets \citep{Wright et al.2011}.
 Its abnormally large radius ($R_p$=1.798 $R_{\rm Jup}$ or 2.037 $R_{\rm Jup}$, depending on the eccentricity)  is difficult to explain even with ohmic heating in its interior \citep{Wu et al.2013, Huang et al.2012}. HAT-P-32Ab orbits a late-type F dwarf at 0.034 AU with a period of 2.15 days \citep{Hartman et al.2011}. The orbit is highly misaligned and lies almost in the same plane as the spin axis of the star ($\lambda=85\degr \pm 1\fdg5$) \citep{Albrecht et al.2012}.  Its host star has a  high radial velocity jitter of 64 $\pm$ 10 m s$^{-1}$ \citep{Knutson et al.2014}. The origin of the host star's high jitter is unclear, although it could be due to  convective inhomogeneities on the stellar surfaces that vary in time, or due to perturbations from an unseen body in the system \citep{Saar et al.1998}.
The  eccentricity of the planetary orbit was poorly constrained because of the high velocity jitter, resulting in two sets of orbital and planetary properties depending strongly on the eccentricity. 
The circular orbit solution is preferred by statistical tests and  the short tidal circularization timescale of the system ($t_{\text{tidal}} \sim3 -5$ Myr, much shorter than the $>$2 Gyr age of the system) \citep{Hartman et al.2011, Zhang et al.2013}. \citet{Seeliger et al.2014} observed 45 transits of HAT-P-32Ab to search for   transit timing variations (TTV), and found no evidence for any perturbations to the hot Jupiter's orbit from a nearby planetary companion.

Using adaptive optics (AO) imaging in the $K_S$ band, \citet{Adams et al.2013} detected a candidate M-dwarf companion at a distance of $\sim$2\farcs9 with a magnitude difference of $\Delta K_S=3.4$, contributing $\sim$4\% of the light in $K_S$ that slightly dilutes the transit signal of the planet. 
{Meanwhile, \citet{Knutson et al.2014} detected a radial velocity trend of $-33\pm10 $m s$^{-1}$yr$^{-1}$ in the system using long-term radial velocimetry, and Ngo et al. (in preparation) confirmed the physically associated stellar companion HAT-P-32B using proper motion measurements from  AO imaging. 
While the velocity jitter might be partially explained by contamination from the M-dwarf, the long-term velocity trend cannot be explained by the  companion star due to its large separation of $\sim$830 AU, but requires an inner body at 3.5-21 AU from the planet host star with a projected mass ($M\sin i$) between 5-500 $M_{Jup}$. }

The hot Jupiter HAT-P-32Ab is very suitable for atmospheric characterization using both secondary eclipses and transmission spectroscopy thanks to its large radius, high temperature, and large atmospheric scale height. \citet{Gibson et al.2013} obtained a low-resolution transmission spectrum in the visible and found a featureless spectrum. The flat spectrum of the planet's terminator can be explained by clouds in the upper atmosphere or a clear atmosphere with  trace amounts of TiO, VO, or metal hydrides that mask the Na and K wings in the spectrum.

In this paper we report measurements  of HAT-P-32Ab's thermal emission spectrum   in the NIR $H$, $K_S$ bands from the ground and the 3.6 and 4.5 \micron~bands from $Spitzer$. To facilitate our characterization of the hot Jupiter's atmosphere, we also obtain high-angular resolution AO imaging of the double star system to characterize both stellar components and to correct for the dilution of the secondary eclipse depths. In Section \ref{obs} we present our observations and data reduction procedures. In Section \ref{analysis} we describe our analysis of the AO images,  the characterization of the M dwarf companion, the analysis of HAT-P-32Ab's secondary eclipse light curves, and corrections to its diluted eclipse depths. We then discuss the eccentricity of the planet's orbit and its atmospheric models in Section \ref{discuss}. Finally, we  summarize our results in Section \ref{summary}.

\section{Observations and data reduction}
\label{obs}

\subsection{Palomar/WIRC Secondary Eclipse Photometry}
\label{wirc}
We observed  two secondary eclipses of the hot Jupiter HAT-P-32Ab in $H$ and $K_S$ bands on UT 2012 October 03 and 2012 October 31 respectively, using the Wide-field Infra-Red Camera (WIRC) at the Palomar 200-in Hale telescope \citep{Wilson2003}. 
The camera had\footnote{The original HAWAII-2 array used for this study failed 
 in April 2014 due to explosive debonding and
separation of the semiconductor from its substrate, and thus is  no longer installed on the camera.} 
a science grade 2048 $\times$ 2048 HAWAII-2 HgCdTe detector with 
 a pixel scale of 0.2487\arcsec/pixel, corresponding to a field of view of $8.7' \times 8.7'$. 
The $H$-band observation started roughly 166 min before the predicted mid-eclipse, and ended   226 min after mid-eclipse. The airmass changed from 1.21 to 1.03 and then back to 1.39 during the observation. To minimize systematics, we ``stared" at the target throughout the observation and used the active guiding scheme developed in \citet{Zhao et al.2012b} to stabilize the telescope motion and keep the stellar centroids at the same positions. We also defocused the telescope to $\sim$2\farcs5  FWHM to  mitigate pixel-to-pixel variations and keep the counts below saturation. 
Due to the  brightness of the target, all images in the $H$ band were taken with  6 sec exposures and one double-correlated sampling (1 Fowler).  
A total number of 1131 images were recorded during the 392 min observing period, corresponding to a duty cycle of 26.3\% when  the readout and centroiding overhead  are taken into account. 
Thirty nine frames had saturated pixels on the target (mostly due to a bright spot on the point spread function caused by astigmatism of the optics) and were thus excluded from subsequent analysis. 
 Forty images with large sudden flux drops due to passing clouds were also excluded. 
 The UTC timestamp of each mid-exposure was converted to the Barycentric Dynamical Time standard (BJD$_{TDB}$) following \citet{Eastman2010}. 

The $K_S$-band observation started 205 minutes before the predicted mid-eclipse and ended 267 minutes after the mid-point, following the same observing strategy as in the $H$-band. The airmass changed from 1.22 to 1.028 before the target transited the meridian, and went back to 1.83 at the end of the observation. We defocused the telescope to $\sim$3\arcsec~  FWHM to keep the counts below saturation. Still, a few images had saturated pixels on the target due to astigmatism and were excluded in the analysis.
We took a total of 1208 non-saturated images with exposures of 8 s exposures during the 472 min observation, corresponding to a duty cycle of 34\%. 

To reduce the images, we constructed and applied darks, twilight flats, and interpolated the bad pixels following the procedures described in \citet{Zhao et al.2012a}. In addition, to better subtract the  background in the science images, particularly the reflected thermal background from the optics, we took dithered sky images immediately before and after the secondary eclipses and averaged them after bias subtraction, flat fielding, and normalization to construct a ``supersky" frame. The background of each science image  was estimated using the normalized ``supersky" and was subtracted after applying the flat field. This new step significantly reduced the large-scale background structures and fringe patterns on the detector, particularly the thermal reflections in the $K_S$ band and the internal fringing in the $H$ band. 

To further reduce detector-related systematics, we implemented a new correction for the non-classical, flux-dependent non-linearity \citep[also known as ``reciprocity failure", ][]{Hill et al.2010, Biesiadzinski et al.2011a, Biesiadzinski et al.2011b} of the HAWAII-2 detector of WIRC. The new calibration improved the precision of the $H$-band data 
whose fluxes suffer from large nonlinearity differences between the wings of the point spread functions (due to low background counts of $\sim$4K) and high fluxes of their peaks ($\gtrsim$31K counts). 
The precision of the $K_S$ band data was not improved by this correction, due to the much smaller difference between the sky background ($\gtrsim$12K counts) and the peak fluxes ($\lesssim$30K) (see Appendix for details).

We corrected for time-varying telluric and instrumental effects in both data sets by selecting
 9 and 12 reference stars with median fluxes between 0.15 - 1.0 times that of HAT-P-32A in the $H$ and $K_S$ band, respectively.
Fainter stars in the field were excluded due to low signal-to-noise, while stars brighter than the target saturate the detector. Stars were also excluded if they caused significantly increased scatter or correlated noise in the residuals in the subsequent light curve analysis. 
{Transient ``hot pixels"    outside the photometry aperture were identified using a local 4-$\sigma$ spatial filter and corrected using a 2-D cubic spline interpolation.}
 We calculated the centroid of each star's position using the flux-weighted average (or ``center-of-light"). 
 The $x$ and $y$ positions of the stellar centroid typically varied by less than  3 pixels in the $H$ band, with a standard deviation of 0.81 pixel in $x$ and 0.52 pixel in $y$. For the $K_S$ band, we had two software glitches that caused the target to drift away by $>$5 pixels. We therefore  excluded the 54 images affected by the glitches from the analysis.  The centroid of the target in the remaining images  fluctuated 
 by less than 3 pixels in both $x$ and $y$,  with a standard deviation of 0.84 pixel in $x$ and 0.60 pixel in $y$, respectively. 

\begin{figure}[t]
\begin{center}
 \includegraphics[width=3.2in, angle=0]{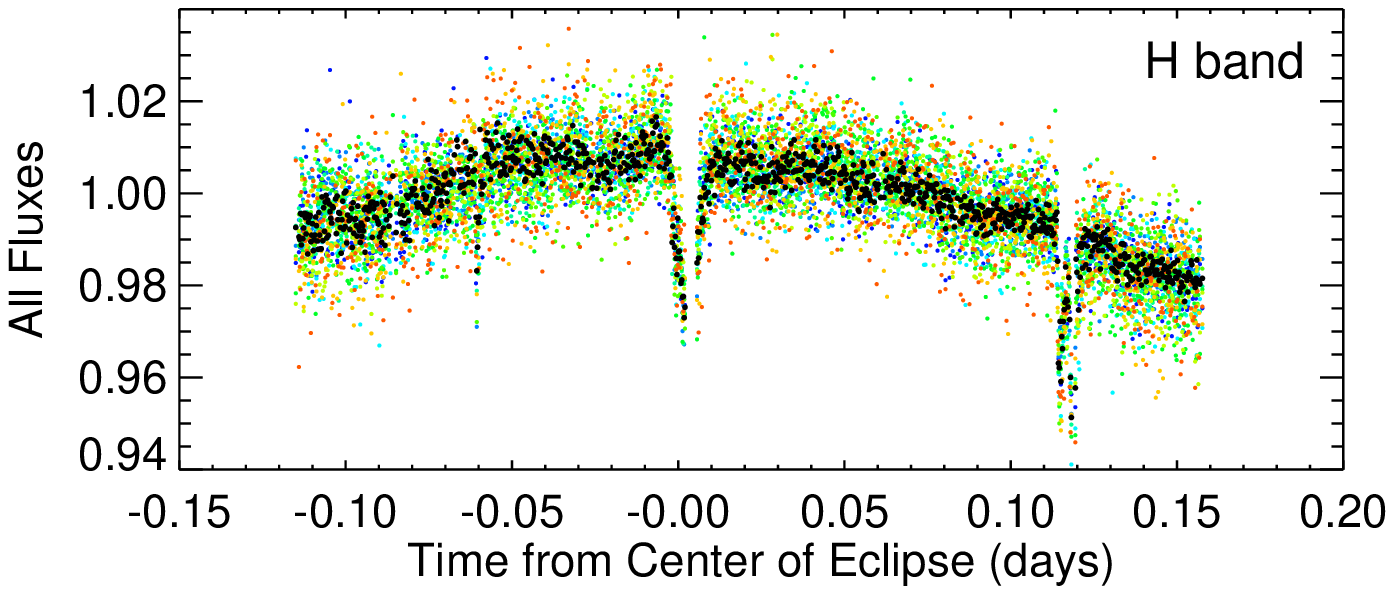}   
 \includegraphics[width=3.2in, angle=0]{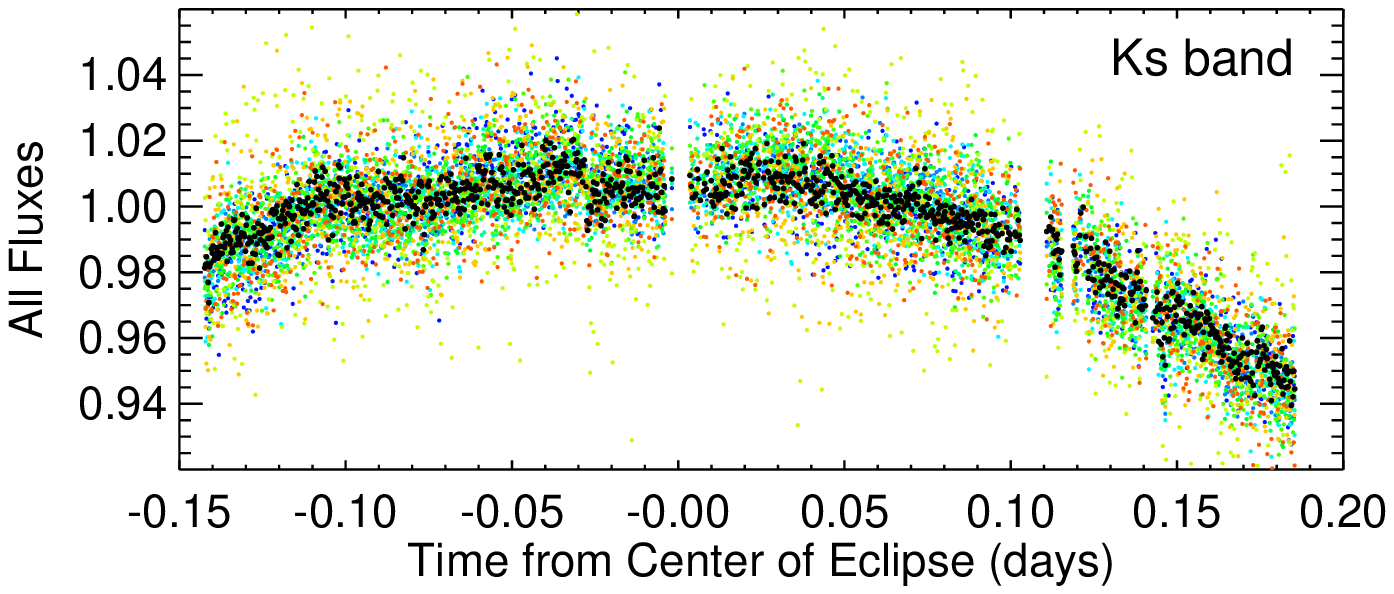}
 \caption{
 Normalized raw light curves of HAT-P-32 in the $H$ band (top) and $K_S$ band (bottom) obtained using Palomar/WIRC. The flux of the target is shown in black dots, while the fluxes of the references stars are shown in colored dots. Data points with large ($>$5\%), sudden flux drops due to passing cirrus (top), or large centroid drifts due to software glitches (bottom) were excluded. 
 Light from the M star companion is also included in our photometric apertures in both bands.
 }
\label{wirc_raw}
\end{center}
\end{figure}

We carried out aperture photometry using elliptical instead of circular apertures to include both the companion M dwarf and the primary star while reducing the encircled background, since the PSFs of the two stars  overlap with each other in the WIRC images. We fixed the position angle of the semi-major axis of the ellipse to 110\degr~based on the analysis of our AO images (section \ref{ao}), and used an axial ratio of 1.4 based on estimation of the contour of the PSFs. 
We applied 48 different aperture sizes with a step of 0.5 pixel for the target and  reference stars. 
The extracted fluxes were normalized to the median of the time series. The median of the reference time series is then taken as the final reference light curve to normalize the flux of HAT-P-32AB to correct for the common-mode systematics stemming from variations of atmospheric transmission, seeing, and airmass, etc.
We found that semi-major axes of 30.5 pixels (7.625\arcsec) and 26.5 pixels (6.625\arcsec) for the $H$ and $K_S$ band, respectively, produced the smallest residuals, and were used as the final photometry apertures. 
We also experimented with circular apertures and time-varying apertures but 
found that these resulted in higher scatter in our final light curve.
 We used elliptical sky annuli with 35-pixel inner and  60-pixel outer semi-major axes in the $H$-band, and 30-pixel inner and 53-pixel outer semi-major axes in the $K_S$-band, respectively, to fit surfaces to estimate and subtract the residual background for the target and references. Different annulus ranges and sizes were also explored but showed consistent results. 
Figure \ref{wirc_raw} shows the normalized raw fluxes of the target and the references in the $H$ and $K_S$ bands, respectively.

\subsection{Spitzer IRAC secondary eclipse photometry}

We observed two secondary eclipses of HAT-P-32Ab with the {\em Spitzer Space Telescope} and the IRAC instrument in the 3.6\micron~and 4.5\micron~band on UT 2011 October 20 and  UT 2011 October 29 respectively. All observations were taken in the full array mode with 256$\times$256 pixels.
We obtained 3972 images with an exposure time of 6 sec in the 3.6 \micron~band and 2167 images with an exposure time of 12 s in the 4.5\micron~band, respectively.  The time stamps in the FITS header, BJD$_{UTC}$, were converted into the Barycentric Dynamical Time standard (BJD$_{TDB}$) or the time of our observations following \citet{Eastman2010}\footnote{BJD$_{TDB}$ $\approx$ BJD$_{UTC}$ + 66.184 s +$\Delta_{TBD-TT}$, where the correction term $\Delta_{TDB-TT}$ is only $\sim$1.6 ms for the two {\em Spitzer} epochs. Unlike the WIRC data, the time stamps in the {\em Spitzer} FITS headers were already in BJD$_{\rm UTC}$.  }.

We extracted photometry from the basic calibrated data (BCD) files generated using version S.19.0.0 of the IRAC pipeline, following the steps described in \citet{ORourke et al.2014}.
Briefly, we first corrected for transient ``hot pixels" within a 20$\times$20-pixel box centered on the target. In total, 0.61\% and 0.21\% of pixels in the 3.6 and 4.5\micron~bands were corrected, respectively. We then calculated the flux-weighted centroid within 3.5 pixels of the approximate position of the target star to find the center of the stellar PSF. The $x$ and $y$ coordinates changed by less than 0.16 and 0.26 pixels, respectively, during our 3.6 and 4.5\micron~observations. 

We experimented with aperture photometry using a time-varying aperture based on the noise-pixel parameter, but we found that fixed photometric apertures gave lower scatter in the final residuals. 
Because the nearby M-dwarf companion and the planet host star are blended together in $Spitzer$ images,
 we used circular apertures with fixed radii of 3.5 pixels in both bands to include the companion while also minimizing errors in the photometry. For the 1.21\arcsec/pixel size of {\em Spitzer IRAC}, our circular aperture corresponds to a radius of 4.24\arcsec, sufficient to encircle the fluxes from both stars with a  separation of 2.93\arcsec (see Section \ref{ao}).
Apertures with radii of 2.7 and 1.8 pixels (3.3\arcsec  and 2.2\arcsec) for our 3.6 and 4.5\micron~band observations, respectively, produced the lowest scatter in the final residuals, but only included partial flux from the companion star and thus were not used in the final solution. We verified that we obtained consistent eclipse depths and time offsets to within 1-$\sigma$ using  apertures ranging from 1.8 to 4.0 pixels. We also verified that, after correcting for the dilution by the companion (see Section \ref{correction}), the eclipse depths obtained with 3.5-pixel apertures became more consistent with the values obtained with 1.8-pixel apertures.
 We estimated the background using the 3-$\sigma$ clipped mean within circular sky annuli with inner and outer radii of 20.0 and 30.0 pixels and 20.0 and 35.0 pixels in the same bands, and obtained consistent values. There were no visible bright stars within these annuli.
 
\begin{figure}[t]
\begin{center}
 \includegraphics[width=3in, angle=0]{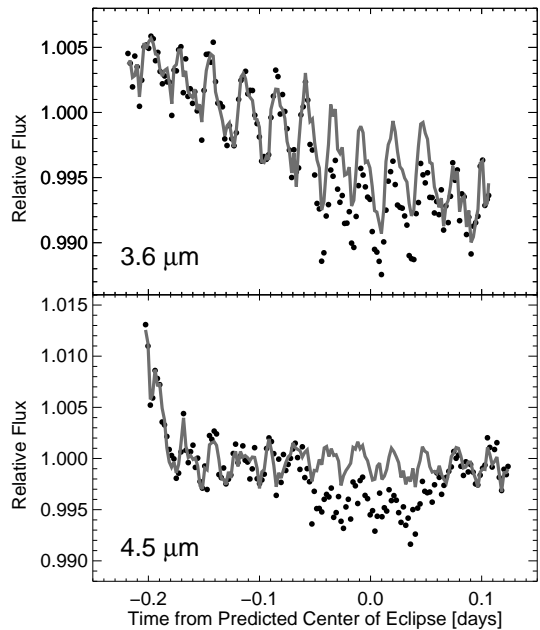}
 \caption{
Normalized raw light curves of HAT-P-32 in the Spitzer 3.6 \micron~and 4.5 \micron~bands. The functions used to correct for intra-pixel sensitivity and linear trends in time are overplotted in grey (Section \ref{spitzer_lc}). Data points are binned in 3 minute intervals. Flux from the faint companion is also included in the photometry.}
\label{spitzer_raw}
\end{center}
\end{figure}

We discarded points in our light curves that suffered uncorrected cosmic ray hits within our photometric aperture or significant spatial drift. In total, we discarded 28 and 23 images from our 3.6 and 4.5 \micron~band light curves, respectively. We also trimmed 9 and 7 frames (corresponding to 0.9 min and 1.4 min) from the beginning of our 3.6 and 4.5\micron~light curves, respectively, in order to avoid effects related to the settling of the telescope at a new pointing. 
Figure \ref{spitzer_raw} shows the normalized raw  light curves of HAT-P-32Ab in both bands.

\subsection{AO Imaging of HAT-P-32AB}

The planet host star HAT-P-32A is in a double star system separated by $\sim2\farcs9$. To better characterize the planet host star  and the planet itself, we carried out AO imaging  to resolve the two stars in both visible and near-IR (NIR) wavelengths.

\subsubsection{Near-Infrared Adaptive Optics Imaging}
We observed HAT-P-32AB in the NIR $H$ and $K_S$ bands on 2013 March 02 using the Keck-II AO system \citep{Wizinowich et al.2000} and the NIRC2 instrument (Instrument
PI: Keith Matthews). 
We used the narrow camera with a scale of  0.01\arcsec/pixel for fine spatial sampling of the point spread function (PSF).
We conducted the observation in position angle mode, in which the orientation of the detector was fixed throughout the observation, instead of using angular differential imaging as the companion was bright enough to be resolved from the primary in this relatively simple observing mode.
We  used the full array with 1024 $\times$ 1024 pixels and the standard three-position dither pattern to assist the subtraction of sky and instrumental background. 
We took a total of 9 images in the $H$ band with 5 s integrations and 24 images in the $K_S$ band with 3 s and 15 s integrations, respectively.
The $K_S$-band data were the same set reported in \citet{Knutson et al.2014}, in which  the images were used to place a limit on the presence of massive outer companions  at smaller projected separations that could explain the observed radial velocity acceleration.
The $H$ and $K_S$ images were also included in Ngo et al.\ (in preparation) as part of their effort to measure the proper motion of both stars.

\begin{figure}[t]
\begin{center}
 \includegraphics[width=3.3in, angle=0]{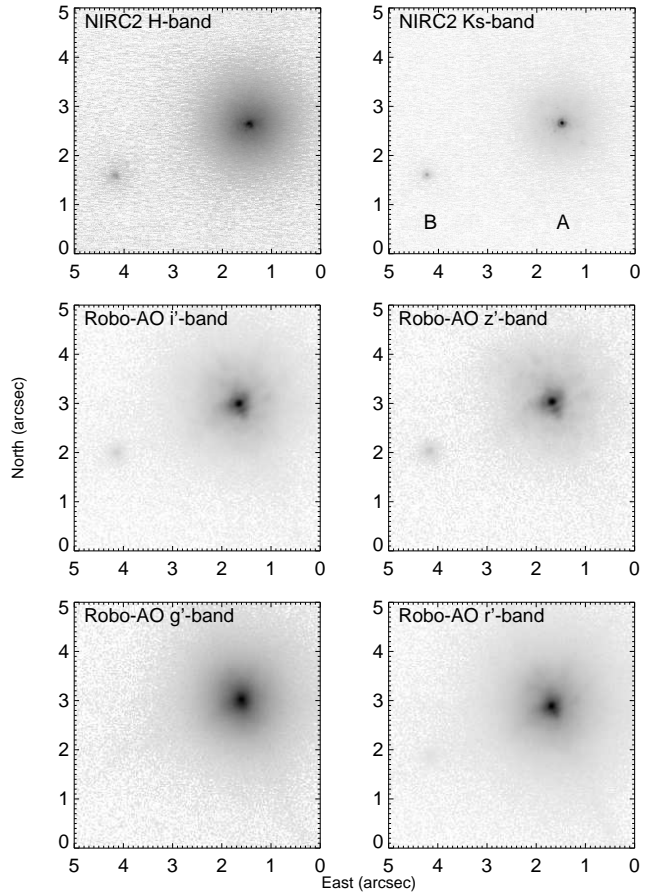}   
 \caption{Adaptive optics imaging of HAT-P-32AB. The binary is well resolved. The F7 primary, HAT-P-32A, is to the right and the fainter companion B is to the left. The top two panels show the Keck-II/NIRC2  images in the near-IR $H$ \& $K_S$ bands. The bottom and middle panels show the Robo-AO images in the SDSS $g', r', i', z'$ bands, respectively. The images are scaled by cubic root to reduce the contrast between the two stars so the faint companion can be easily seen in the figure. The images are rotated such that north is up and east is to the left. }
\label{aoimg}
\end{center}
\end{figure}

After  initial dark subtraction and flat-field correction using dome flats,  we subtracted the averaged sky and instrumental background in each science frame using dithered images at different positions. This worked well as the three dither positions fell on different quadrants on the detector. Figure \ref{aoimg} shows two representative images of the HAT-P-32AB system obtained with NIRC2 in $H$ and $K_S$. 
 We treated each science image individually in the subsequent analysis instead of  aligning and co-adding them together as the seeing and PSFs varied significantly from frame to frame. One dither position fell on the lower-left quadrant of the detector with higher bias and read noise level, while another dither position fell too close to the edge of the detector, causing the wing of the companion's PSF only partially imaged on the detector.

\subsubsection{Visible Adaptive Optics Imaging}
We observed HAT-P-32AB at visible wavelengths on UT 2013 January 20 using the Robo-AO instrument, an autonomous laser-guide-star adaptive-optics imaging system installed on the 60-inch telescope at the Palomar Observatory \citep{Baranec et al.2013, Baranec et al.2014}. 
{Robo-AO provides  diffraction-limited full-width-at-half-maximum (FWHM) of $0\farcs10 - 0\farcs15$  and Strehl ratios of 4-26\% in the $i'$-band with a pixel sampling of 0\farcs04353/pixel \citep{Baranec et al.2014}, sufficient to resolve the stellar companion from the primary while providing fine sampling of the PSF. }
 The observation consisted of a sequence of rapid-frame-transfer read-outs at 8.6 frames per second with a total integration time of 90 s in the   Sloan Digital Sky Survey (SDSS) $g', r', i', z'$ bands, respectively. 

The images were reduced using the pipeline described in \citet{Law et al.2014}. 
{In short, after dark subtraction and flat-fielding using daytime calibrations, the individual images were up-sampled, and then shifted and aligned by cross-correlating with a diffraction-limited PSF. The aligned images were then co-added together using the Drizzle algorithm \citep{Fruchter et al.2002} to form a single output frame for each bandpass. The final ``drizzled" images have a finer pixel scale of 0\farcs02177/pixel.}
Figure \ref{aoimg} shows the images of  HAT-P-32AB obtained with Robo-AO in the $g'r'i'z'$ bands.

\section{Analysis and Results}
\label{analysis}

\subsection{Analysis of AO images}
\label{ao}

\subsubsection{PSF modeling}
\label{psf}
We first analyzed the AO images of HAT-P-32AB to determine their separation, position angle, and flux ratios in $H$ and $K_S$ bands. 
Given the well separated PSFs of the two stars in the NIRC2 images, we started with standard aperture photometry of the two stars.  However, due to residual instrumental background patterns on the InSb array of NIRC2,  the varying and high level of uncorrected bias patterns in the lower-left quadrant of the detector where one dither position fell on, and a partially imaged PSF wing in the third dither position, we were unable to obtain reliable flux ratio measurements. Instead, we determined the flux ratios using PSF model fitting.  Nonetheless, the separation and position angle can still be robustly determined using aperture photometry and they are consistent with those determined from PSF modeling within 1-$\sigma$.

We constructed PSF models for both stars simultaneously using a joint Gaussian and Moffat PSF function, following a similar approach to the one used in \citet{Bechter2014}. The Gaussian function is used to characterize the core of the PSF, while the Moffat function is used to trace the extended PSF wing (or halo). This joint function can model the effects of tip/tilt and focal anisoplanatism well, although it cannot account for higher order aberrations or diffraction in the PSF. Nonetheless, it is sufficiently accurate for our purpose here to determine the relative flux ratio,  separation, and  position angle of the two stars \citep{Bechter2014}. 
We assumed the same PSFs for the two stars and only allowed their flux ratio to change, which is justified by their close angular distance ($\sim$2\farcs9) and
corresponding high degree of similarity in their PSFs. 
The PSF model has a total number of 15 free parameters, including the flux ratio, background, the peaks of the Gaussian and Moffat profiles, the FWHMs along the $x$ \& $y$ axes of the detector, the $x$ \& $y$ centroid positions, and the individual position angles of the Gaussian and Moffat profile.   

To test the reliability of this model, we simulated NIRC2 and Robo-AO images using the actual background and the average PSF model, and injected a scaled version of the PSF as the companion at certain separations and position angles. We then fitted the simulated data using the Levenberg-Marquardt algorithm for  least-square minimization, and  explored the parameter space extensively. 
{We demonstrate that we were able to recover the injected flux ratios within an accuracy of 5-10\% for the NIRC2 data. For the Robo-AO data, we can recover the flux ratios with an accuracy better than 30\% at separations larger than 2\farcs5, due to more extended PSF halo and much fainter flux from the companion.} 
We thus proceeded to implement this modeling approach to the NIRC2 and Robo-AO images.  

\subsubsection{Application to NIRC2 and Robo-AO images}

\begin{figure}[t]
\begin{center}
 \includegraphics[width=3in, angle=0]{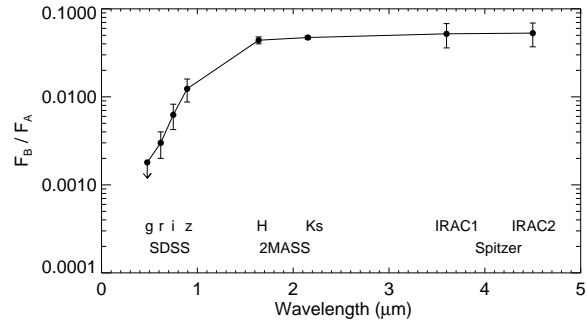}   
 \caption{Flux ratios of the fainter companion HAT-P-32B to the brighter planet host HAT-P-32A vs. wavelengths. Flux ratios in the SDSS bands and the 2MASS bands are determined using AO imaging and PSF fitting. Flux ratios in the $Spitzer$ 3.6 and 4.5 \micron~bands are extrapolated using the $K_S - [3.6]$ and $K_S - [4.5]$ colors determined from SED fitting. 
 }
 \label{stellar_ratio}
\end{center}
\end{figure}

For the NIRC2 images, we trimmed the reduced images into a smaller size of 600$\times$200 pixels, which includes both stars while avoiding extra background.
 We fitted the double PSF model to each of the 9 images in $H$ band and 24 images in $K_S$ band.  The pixels in each trimmed image were weighted by their photon noise or background noise, whichever is higher. 
Since the weights were dominated by high signal-to-noise pixels in the core of the PSF, this model fitting approach allows us to obtain reliable  parameters even for images affected by variable and high bias levels (the lower left quadrant) and truncated PSF wings.
We took the error-weighted average of the  best-fit parameters for each set of images as the final result. The scatters of the best-fit parameters dominate their formal uncertainties from the fits and were thus chosen as the final uncertainties of the parameters. 
The resulting flux ratios (${f_B}/{f_A}$) for $H$ and $K_S$ bands are 0.044 $\pm$ 0.005 and 0.047 $\pm$ 0.002, respectively, 
and are consistent with those from Ngo et al. (in preparation) within 2-$\sigma$.
These values are plotted in Figure \ref{stellar_ratio} and are shown in Table \ref{hat32ab}, together with the corresponding apparent magnitudes for each star, where the individual magnitudes are derived using the total magnitudes of the system from the tenth data release of SDSS \citep{Ahn et al.2014}.

We convert the separation of HAT-P-32AB from pixels to arcseconds using the plate scale and orientation of the NIRC2 array determined by \citet{Yelda2010}\footnote{Pixel scale = 9.950 $\pm$ 0.004 mas/pixel. Actual Position Angle = Measured Position Angle $-0\fdg254 \pm 0\fdg016$ \citep{Yelda2010}.}. 
The resulting best-fit separation $\rho$ and position angle $\theta$ in the $H$-band are: $\rho=2\farcs927 \pm 0\farcs011$ and $\theta=110\fdg65 \pm 0\fdg18$.
The $K_s$ band NIRC2 images gave highly consistent results of $\rho=2\farcs925 \pm 0\farcs006$ and $\theta=110\fdg64 \pm 0\fdg15$. The weighted averages of the separation and position angle are listed in Table \ref{hat32ab}.

As another cross-check of the NIRC2 results, we also selected the best Palomar/WIRC images of HAT-P-32AB with well-focused and well-separated PSFs  in both bands obtained when checking the telescope focus, and fitted for the flux ratios.  The resultant flux ratios and their formal errors are 0.040 $\pm$ 0.006 in $H$ and 0.058 $\pm$ 0.011 in $K_S$, and the separation is  2\farcs917 $\pm$ 0\farcs036, consistent with the NIRC2 results within 1-$\sigma$. The position angle is not directly comparable as the orientation of the WIRC detector is uncalibrated. 

The Robo-AO images were analyzed in the same manner using the dual two-component PSF model. We trimmed the reduced Robo-AO images into a smaller size of 610$\times$420 pixels to avoid extra background in the fit. The best-fit reduced $\chi^2$ ranges from 0.22 to 0.26 for the four SDSS bands. To avoid possible underestimation from the nominal uncertainties from the best-fit, we chose a conservative uncertainty of 30\% of each flux ratio in the four bands based on our simulations described previously (see Section \ref{psf}). We detected the companion star in the $r', i',$ and $z'$ bands, and derive an upper limit for the flux ratio in the $g'$ band. The best-fit flux ratios are listed in Table \ref{hat32ab} and are shown in Figure \ref{stellar_ratio}. 
We also calibrated the positions of the two stars on the detector  as an additional check on the NIRC2 results. 
The resulting error-weighted average separation and position angle from the three bands  with a detection of the companion are 2\farcs920 $\pm$ 0\farcs006 and 111\fdg50 $\pm$ 0\fdg32, and are consistent with the NIRC2 results at the 1-$\sigma$ and 3-$\sigma$ level, respectively, where the uncertainties are dominated by calibration of the instrument's pixel scale and orientation.
We take the error-weighted average of the NIRC2 and Robo-AO separations as the final separation of the binary. We  take the NIRC2 position angle as the final solution thanks to its better calibration, higher angular resolution and finer pixel scale. These results are listed in Table \ref{hat32ab}.

\begin{deluxetable}{lccc}
\tabletypesize{\scriptsize}
\tablecaption{Properties of HAT-P-32AB}
\tablehead{
\multicolumn{4}{c}{Apparent magnitudes}\\
&		\colhead{AB combined}				&\colhead{Star A\tablenotemark{a}} & \colhead{Star B\tablenotemark{a}} 
} 
\startdata
 ${g'}$		&		 11.482$\pm$0.001\tablenotemark{b}	& \nodata  & \nodata	\\
 ${r'}$		&		 11.167$\pm$0.001\tablenotemark{b}	&	11.170$\pm$0.001	& 17.477$\pm$0.362		\\
 ${i'}$		&		 11.062$\pm$0.001\tablenotemark{b}	&	11.069$\pm$0.002	& 16.579$\pm$0.347		\\
 ${z'}$		&		 11.422$\pm$0.003\tablenotemark{b}	&	11.435$\pm$0.004	& 16.207$\pm$0.317		\\
$H$				&		 10.024$\pm$0.022\tablenotemark{c}	&	10.071$\pm$0.022	& 13.462$\pm$0.101	 \\
$K_S$			&          \phn9.990$\pm$0.022\tablenotemark{c}		&	10.040$\pm$0.022 	& 13.355$\pm$0.051	\\
\cutinhead{B to A flux ratios }
$(f_B/f_A)_{g'}$		&	\multicolumn{3}{c}{ $<$0.0018 		}		\\	
$(f_B/f_A)_{r'}$			&	\multicolumn{3}{c}{ 0.003$\pm$0.001	}			\\
$(f_B/f_A)_{i'}$			&	\multicolumn{3}{c}{ 0.006$\pm$0.002		}		\\
$(f_B/f_A)_{z'}$			&	\multicolumn{3}{c}{ 0.012$\pm$0.004		}		\\
$(f_B/f_A)_{H}$			&	\multicolumn{3}{c}{ 0.044$\pm$0.005		}	 	\\
$(f_B/f_A)_{K_S}$		&	\multicolumn{3}{c}{ 0.047$\pm$0.002		}		\\
$(f_B/f_A)_{3.6}$	&	\multicolumn{3}{c}{ 0.050$\pm$0.020		}		\\
$(f_B/f_A)_{4.5}$	&	\multicolumn{3}{c}{ 0.053$\pm$0.020		}		\\
\cutinhead{Effective temperatures from SED fit}
$T_{\text{eff, A}}$ (K)	&	\multicolumn{3}{c}{6269 $\pm$ 64 }				 \\
$T_{\text{eff, B}}$ (K)	&	\multicolumn{3}{c}{3565 $\pm$ 82 }			 \\
\cutinhead{AO Astrometry}
					&      NIRC2	& Robo-AO & Final result\tablenotemark{d} \\
Separation			&	{2\farcs925 $\pm$ 0\farcs005 } 	& 2\farcs920 $\pm$ 0\farcs006	&  2\farcs923 $\pm$ 0\farcs004 \\
P.A. ($\vec{{AB}}$)		 &	 110\fdg64 $\pm$ 0\fdg12	& 111\fdg50 $\pm$ 0\fdg32  &  110\fdg64 $\pm$ 0\fdg12  	 
 \enddata
 \tablenotetext{a.}{Derived using total magnitudes and flux ratios in this work.}
 \tablenotetext{b.}{Magnitude from the tenth data release of SDSS \citep{Ahn et al.2014}.}
 \tablenotetext{c.}{2MASS magnitude}
 \tablenotetext{d.}{We adopt the position angle from NIRC2 as the final nominal solution because of its better instrumental calibration, higher angular resolution and finer pixel scales.}
\label{hat32ab}
\end{deluxetable}

\vspace{0.3in}

\subsection{Stellar SEDs}

\begin{figure}[t]
\begin{center}
\includegraphics[width=3in, angle=0]{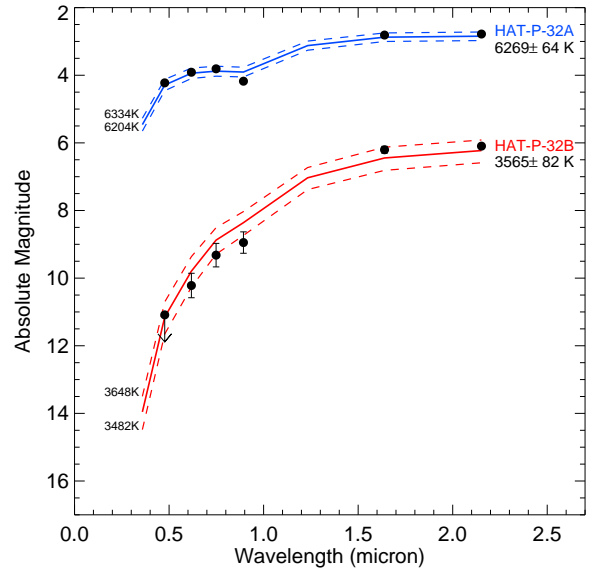}   
 \caption{Best-fit SED models for HAT-P-32AB based on synthetic magnitudes of  PHOENIX model spectra in the observed bands.
 Filled dots with error bars show absolute magnitudes for each star derived using their flux ratios and total magnitudes (some error bars are too small to be seen).
  The solid lines show the best-fit models.  The dashed lines show their corresponding 1-$\sigma$ uncertainties with $T_{\rm eff}$ values  labeled to the left.
The absolute magnitudes are calculated with a distance modulus of 7.259 ($d=283$ pc) based on \citet{Hartman et al.2011}.
}
\label{sed}
\end{center}
\end{figure}

In order to account for the dilution of our measured secondary eclipse depths in the {\em Spitzer} 3.6 and 4.5 \micron~bands, we must first estimate the flux ratios of the binary in these two bands.  We do this by fitting the Spectral Energy Distributions (SEDs) of the two stars using the synthetic broadband magnitudes converted from PHOENIX  model spectra by the Dartmouth Stellar Evolution models \citep{Dotter et al.2008, Husser2013}. Specifically, we converted the combined magnitudes of  HAT-P-32AB into individual absolute magnitudes based on their measured flux ratios and the distance $d= 283$ pc \citep{Hartman et al.2011}, and fitted the SEDs using a Dartmouth model with [Fe/H] = -0.04 and [$\alpha$/Fe]=0 \citep{Hartman et al.2011}. We excluded the upper limit on the $g'$ band magnitude of HAT-P-32B in the fit. 
We then fit for the effective temperatures of both stars, where we take into account the uncertainty in their metallicity by fitting SED models for the 1-$\sigma$ lower and upper limits on [Fe/H] from a spectroscopic analysis of the F star primary.
The resulting systematic differences among models are added in quadrature to the best-fit uncertainty of \teff.
The best-fit SED models for HAT-P-32AB are shown in Figure \ref{sed}. The best model for HAT-P-32B has a temperature of 3565 $\pm$ 82 K, indicating a spectral type of  M1.5 according to \citet{Kraus et al.2007} and \citet{Lepine et al.2013}.  The best model for HAT-P-32A gives  \teff~= 6269 $\pm$ 64 K, consistent with the value of \citet{Hartman et al.2011} within 1-$\sigma$.
We derived the flux ratios of HAT-P-32AB  in the {\em Spitzer} bands based on the $K_S$ - [3.6] and $K_S$ - [4.5] colors from the SED models and listed them in Table \ref{hat32ab}. The $K_S$ - [3.6] and $K_S$ - [4.5] colors from the models are consistent with the empirical values in \citet{Patten et al.2006}.
Uncertainties of the derived flux ratios were propagated from the magnitudes of the SED models in the two {\em Spitzer} bands.  Figure \ref{stellar_ratio} also shows  the flux ratios of HAT-P-32AB in the {\em Spitzer} bands.

\subsection{Analysis of Spitzer Light Curves}
\label{spitzer_lc}
 
We simultaneously fit our light curves with secondary eclipse models \citep{Mandel2002}  and decorrelation functions that
 correct for the well-known intra-pixel sensitivity effect. For both observations, we use the decorrelation function,
\begin{equation}
F(\{c_i\},\bar x,\bar y,t)=c_0 + c_1\bar x + c_2 \bar y + c_3 \bar x^2 + c_4 \bar y^2 + c_5 t, \label{eq:dec}
\end{equation}
where the $\bar x$ and $\bar y$ are the median-subtracted centroid positions, $t$ is time from the predicted center of secondary eclipse assuming a circular orbit, and the $\{c_i\}$ are free parameters. We use the Bayesian Information Criterion (BIC) to verify that we achieve optimal results by including all of the terms in
Eq.~\ref{eq:dec} \citep{Liddle2007, Priestley1981}. The BIC penalizes extra model parameters, and models with lower BIC are usually preferred.
Specifically, for the 4.5~$\mu$m band observation, the function with all of the terms gives the lowest BIC compared to a function without the quadratic terms ($\Delta BIC=-2$)\footnote{$|\Delta BIC| > 1$ means significant difference, and the model with lower BIC is preferred \citep{Priestley1981}.}.
Similarly, for fits to the 3.6~$\mu$m band data, the function with all of the terms gives the lowest BIC compared to that without the quadratic terms ($\Delta BIC=-63$).
We tried adding a cross-term ($c_6 \bar{x} \bar{y}$) to each decorrelation function, but this did not improve our fits for either bands.

\begin{figure}[t]
\begin{center}
 \includegraphics[width=3in, angle=0]{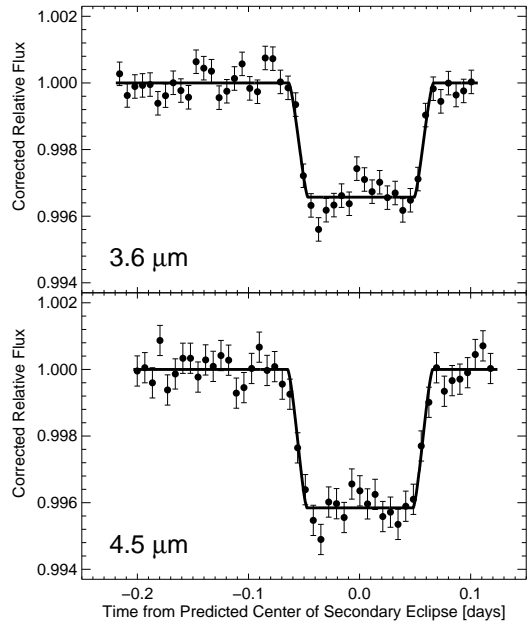}   
 \caption{The final secondary eclipse light curves of HAT-P-32Ab in the {\em Spitzer} 3.6 \micron~and 4.5 \micron~bands after correcting for intra-pixel sensitivity variations. Best-fit light curve models are over plotted as solid lines. 
 Data are binned in 10 minute intervals. The sizes of the error bars are binned based on the RMS scatter (0.327\% and 0.324\% for the top and bottom panels, respectively) in the residuals between the photometry and the best-fit eclipse model over the whole light curve. 
 Note that the secondary eclipse depths are diluted by the M dwarf stellar companion. }
\label{spitzer_lc_fig}
\end{center}
\end{figure}

\begin{figure}[t]
\begin{center}
 \includegraphics[width=3in, angle=0]{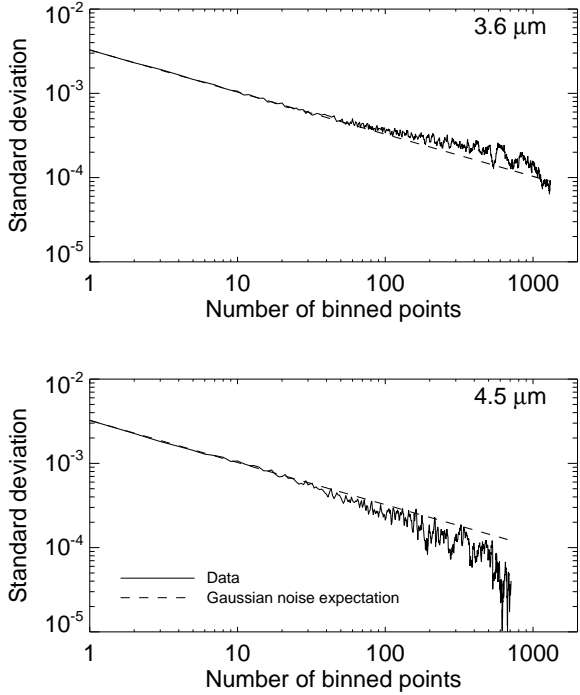}   
 \caption{Comparison of HAT-P-32Ab's residual scatter with Gaussian noise expectation for the {\em Spitzer} bands. The dashed lines show the standard deviation of the residuals if they follow a Gaussian (or white noise) distribution. The solid lines show the standard deviation of actual residuals as a function of bin size. 
 Both curves closely follow the ``white-noise" expectation with bin sizes smaller than 50 points ($\sim$6 minutes for the top and $\sim$10 min for the bottom), indicating that most of the small-scale intra-pixel variations have been corrected. 
}
\label{spitzer_scatter}
\end{center}
\end{figure}

We first perform our simultaneous fit using the Levenberg-Marquardt algorithm \citep{Press1992}. We report the resulting values for the eclipse depths and offsets as our best-fit results. The decorrelation functions for the raw relative flux are shown in Figure  \ref{spitzer_raw}. 
Figure \ref{spitzer_lc_fig} shows the final secondary eclipse light curves and best fit models for HAT-P-32Ab after removing the correlated systematics. 
We assume a constant error for each measurement of relative flux, equal to the root-mean-squared (RMS) scatter in the residuals between our best-fit eclipse model and decorrelated data, as it better represents the uncertainty in the relative flux than the nominal photon noise.
The RMS scatters in our final residuals are 0.327\% and 0.324\% in the 3.6\micron~and 4.5\micron~bands, respectively, which are 23.6\% and 26.9\% above the photon noise limits. 
Figure \ref{spitzer_scatter} compares the residual scatter with ``white noise" expectation, and indicates that the standard deviation of the residuals roughly follow a Gaussian distribution when binned together, suggesting that the decorrelation functions have removed most of the systematics. 

We  estimate our uncertainties using two different  methods. First, we use the Markov-Chain Monte Carlo (MCMC) method with 10$^7$ steps and compute the width of the 68.3\% (1-$\sigma$) symmetric confidence intervals centered on the medians of our (roughly Gaussian, uncorrelated) parameter distributions. We also use the residual permutation (RP) method \citep{Winn et al.2009} to provide an estimate of errors accounting for any time-correlated noise in our data. We report the larger of the two as our formal errors. 

For the 3.6\micron~band data, the RP and MCMC errors on the diluted eclipse depth and offset are 0.010\% and 0.014\% and 1.4 min and 1.3 min, respectively. Thus, we measure a diluted depth of 0.341\% $\pm$ 0.014\% and an offset of 1.3 min $\pm$ 1.4 min. For the 4.5\micron~data, the RP and MCMC errors on the diluted eclipse depth and offset are 0.005\% and  0.9 min and 0.018\% and 1.7 min, respectively. Thus, we measure a diluted depth of 0.418\% $\pm$ 0.018\% and a delay of -1.2 min $\pm$ 1.7 min in this band. 
We list these results in Table \ref{tab3}.
 Our observed center of eclipse times for the 3.6 and 4.5 \micron~bands
are $T_{se}=2455855.57877 \pm 0.00095$ (BJD$_{TDB}$) and 2455864.17701 $\pm$ 0.00118 (BJD$_{TDB}$), respectively.

\subsection{Analysis of Palomar/WIRC light curves}
\label{wirc_lc}

The $H$ and $K_S$ band Palomar light curves of HAT-P-32Ab do not show strong correlated systematics after normalizing with the median reference light curves, but still need further decorrelation to detect the secondary eclipse signal.
We fit a secondary eclipse light curve model simultaneously with a decorrelation function to the data,  
\begin{equation}
f(\{ a_i \}, t) = a_0 + a_1t + a_2t^2 +  a_3 L_{\text{Ref}}, 
\end{equation}
where $f$ is the reference-corrected flux, $t$ is time from the predicted center of secondary eclipse assuming a circular orbit, and the $\{a_i\}$ are free parameters. 
Similar to \citet{ORourke et al.2014}, we also included the median reference light curve $L_{\text{Ref}}$ in the model as it further reduces the scatter of light curves in both bands. 
The scatter in the target's centroid is relatively small, and we do not see obvious correlation between the light curve flux and  centroid positions.  We therefore do not include the centroid positions in the decorrelation function, similar to that in \citet{Zhao et al.2012b}. 
In addition to the decorrelation coefficients, the secondary eclipse depth is the only free parameter in the fit. 
The orbital ephemeris  is fixed to the circular  solution of \citet{Hartman et al.2011}, i.e., $T_0 = 2454420.44637$ (BJD) and Period =  2.150008 days,  as our $H$ or $K_S$ data cannot constrain the eclipse timing with better precision than that of the $Spitzer$ data. The eclipse duration, inclination, semi-major axis, and stellar and planetary radii are also fixed to the circular orbit solution  of \citet{Hartman et al.2011} based on our $Spitzer$ secondary eclipse timing (see Table \ref{tab2} and Section \ref{ecc}).

\begin{figure}[t]
\begin{center}
 \includegraphics[width=3in, angle=0]{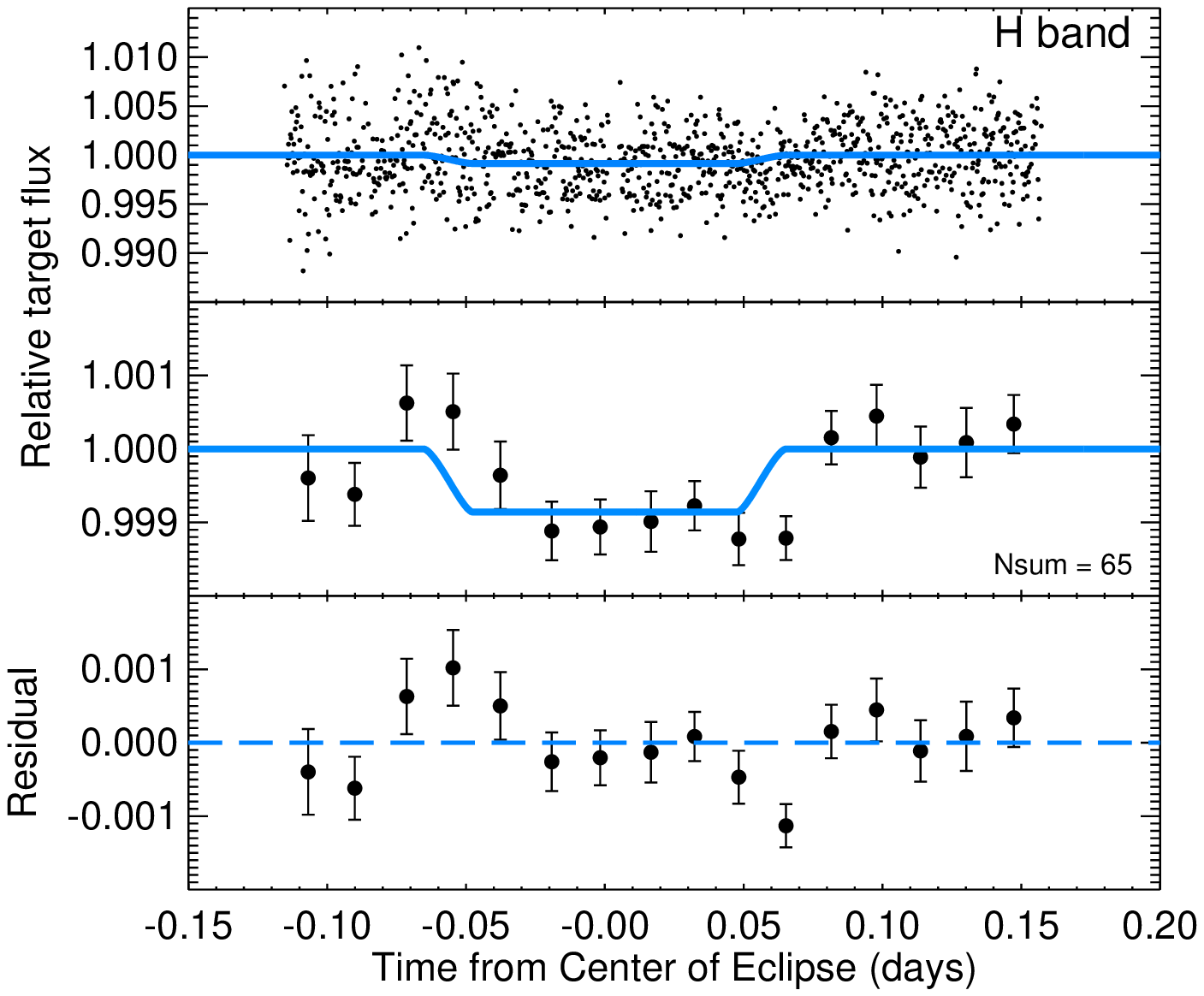}   
 \includegraphics[width=3in, angle=0]{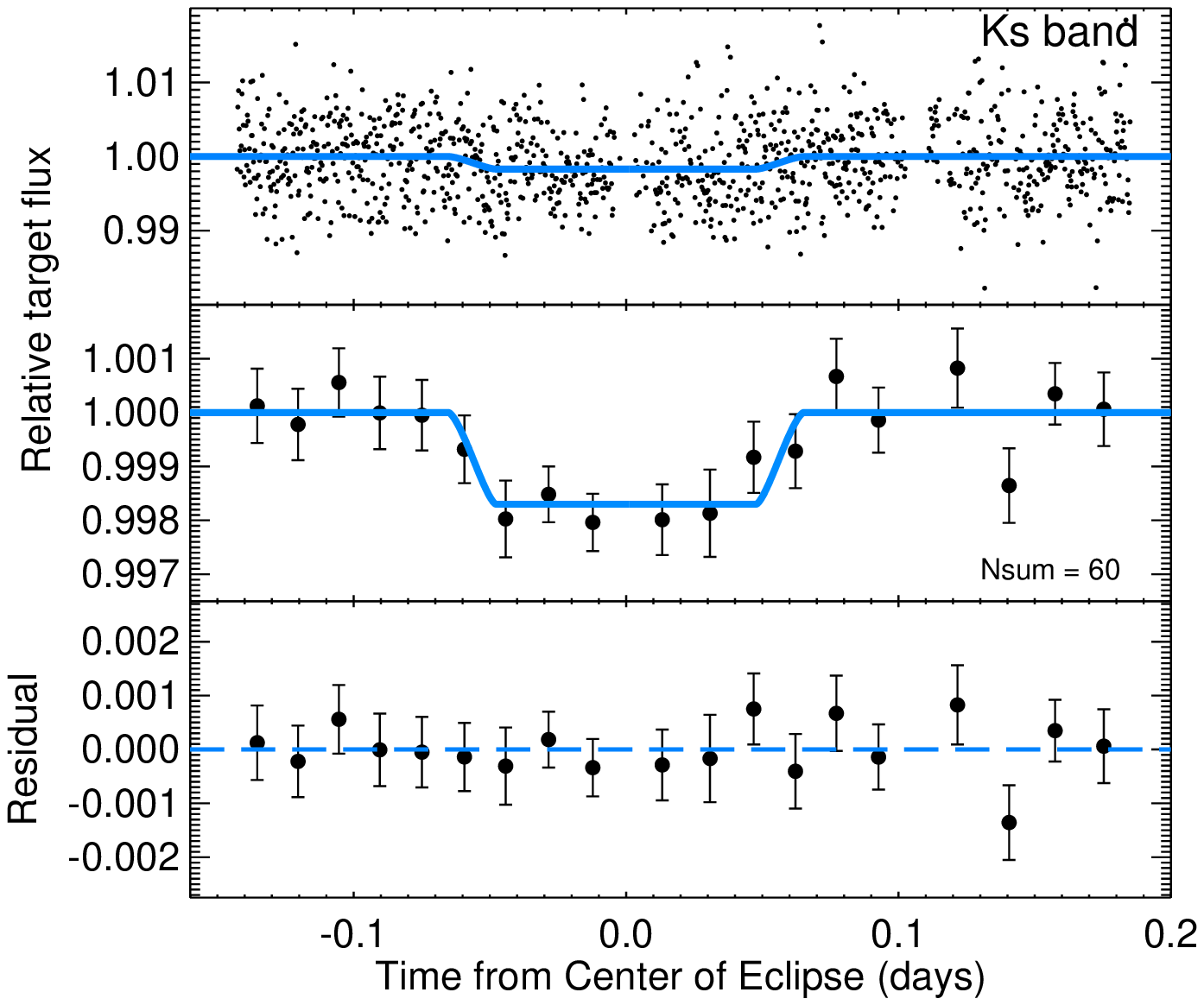}
 \caption{The final decorrelated secondary eclipse light curves of HAT-P-32Ab obtained using WIRC. Best-fit light curve models are over plotted as solid blue lines. The top panels show the unbinned, normalized data after decorrelation. The middle panels show the data with 23-minute bins, while the bottom panels show the corresponding residual after subtracting the best-fit models. The sizes of the error bars are based on the scatter of the data in each bin. Note that the secondary eclipse depths are diluted by the stellar companion.  
  }
\label{lc}
\end{center}
\end{figure}

\begin{figure}[t]
\begin{center}
 \includegraphics[width=3in, angle=0]{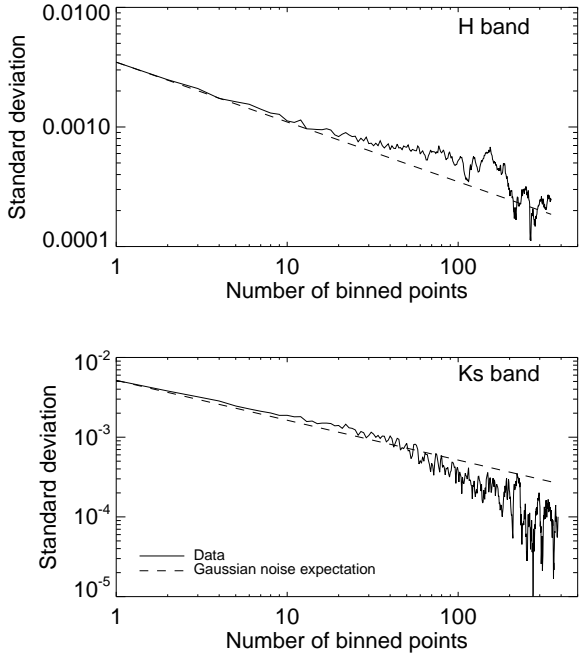}   
 \caption{Comparison of HAT-P-32Ab's residual scatter with Gaussian noise expectation in the $H$ and $K_S$ bands. The dashed lines show the standard deviation of the residuals if they follow a Gaussian distribution (or white noise). The solid lines show the standard deviation of actual residuals as a function of bin size. The residuals are not following the Gaussian expectation in the $H$ band beyond bin size of 20 points, indicating the presence of time-correlated ``red-noise" in the data.}
\label{wirc_resi}
\end{center}
\end{figure}

We employed the Levenberg-Marquardt algorithm in order to determine the best-fit solution. We searched the parameter space extensively with a fine grid of starting points to ensure that we  find the global minimum instead of local minima. 
The data points are uniformly weighted such that the reduced $\chi^2$ is nearly 1.0.  
We also tested the necessity of each term in our decorrelation function using the  Bayesian Information Criterion. For the $K_S$ band data set, a decorrelation function with all  coefficients ($a_0$ to $a_3$) gives the lowest BIC value ($|\Delta BIC|>$1)  and thus is preferred for the final fits. 
For the $H$ band data, a decorrelation function without the quadratic term $a_2$ gives a significantly lower BIC value than other models ($|\Delta BIC|>$6); therefore only $a_0, a_1$, and $a_3$ are used for the final fits.  
The  global best-fit solution for the $K_S$ band data gives a diluted eclipse depth of 0.170\% $\pm$ 0.035\% with a reduced $\chi^2$ of 0.96. For the $H$ band data, the best-fit diluted depth is 
 0.086\% $\pm$ 0.024\% with a reduced $\chi^2$ of 0.92. The best-fit light curve models are shown in the bottom panels of Figure \ref{lc}.
 Figure \ref{wirc_resi} compares the scatter of the best-fit residuals with ``white noise" expectations, indicating there are still  time-correlated systematics in both light curves, particularly in the $H$-band.

We  verify the robustness of the best-fit eclipse depths and estimate their uncertainties using two methods:  bootstrapping \citep{Press1992} and  residual permutation \citep{Winn et al.2009}. For the bootstrapping method, we uniformly resample the data with replacement and re-fit the light curve with the aforementioned decorrelation model. This technique is suitable for unknown distributions, and can robustly test the best-fit model and the distribution of the parameters. We made 2000 bootstrap iterations and the resulting distribution for the eclipse depth is nearly Gaussian. The corresponding 1-$\sigma$ uncertainties are 0.017\% in $H$ and 0.054\% in $K_S$, respectively. For the residual permutation method, 
we subtract the best-fit model from the data and shift the residuals pixel-by-pixel. The shifted  residuals are then added back to the best-fit model and are re-fitted. We conducted 2184 iterations in the $H$ band and 2108 iterations in the $K_S$ band. The resulting median diluted eclipse depth and 1-$\sigma$ uncertainty is 0.085\% $\pm$ 0.032\% in the $H$ band,  and 0.171$^{+0.031}_{-0.043}$\% in the $K_S$ band. 
Our two estimates of the secondary eclipse depth and corresponding uncertainty are consistent with each other to well within 1-$\sigma$.  
We report the best-fit eclipse depths as the final result in Table \ref{tab3}. We compare the uncertainties from all the methods and report the largest value as our formal errors.

\subsection{Flux ratio correction}
\label{correction}
Due to the close angular separation of the planet host star and the M1.5 companion,  our  {\em Spitzer} and WIRC photometry included both stars in their apertures.  The measured secondary eclipse depths of the hot Jupiter are therefore diluted by the flux from the companion star. The  true planet-to-star flux ratio is 
\begin{equation}
\frac{f_b}{f_A} = \delta \left(1+\frac{f_B}{f_A} \right), 
\end{equation}
where $\delta$ is the measured, diluted eclipse depth, $\frac{f_B}{f_A}$ is the flux ratio of the M dwarf companion HAT-P-32B to the planet host star HAT-P-32A (see Table \ref{hat32ab}), and (1+$\frac{f_B}{f_A})$ is the dilution correction factor \citep[see also][]{Shporer et al.2014}. We calculated the dilution correction factors  based on the measured and derived flux ratios in Table \ref{hat32ab}, and applied them to  correct for the four secondary eclipse bands respectively, i.e., $H$, $K_S$, IRAC 3.6 \micron~and 4.5 \micron. We propagated the errors to calculate the final uncertainties. The final dilution-corrected planet-to-star flux ratios are listed in Table \ref{tab3}.  

This dilution will also affect estimates of HAT-P-32Ab's transit depth, depending on the photometry apertures used and the actual amount of flux contamination included in the analysis. 
Existing transit observations for this planet were typically obtained at visible wavelengths, where the flux ratio of the M dwarf companion should be relatively small as compared to our infrared data.
 \citet{Gibson et al.2013} took the M dwarf  into account in their transit light curve analysis. They  found minimal contaminations in their light curves
 as a result of their small apertures, and obtained transit parameters that were consistent with those of \citet{Hartman et al.2011}.

\begin{deluxetable*}{lcccc}[th]
\tabletypesize{\scriptsize}
\tablecaption{HAT-P-32Ab secondary eclipse parameters}
\tablehead{
\colhead{Parameter} & \colhead{$H$} & \colhead{$K_S$} & \colhead{3.6\micron} &\colhead{4.5\micron} 
} 
\startdata
Diluted eclipse depth 		  & 0.086\% $\pm$ 0.032\%	& 0.170\% $\pm$ 0.054\% & 0.341\% $\pm$ 0.014\%	& 0.418\% $\pm$ 0.018\%  \\
Dilution corrected eclipse depth  & 0.090\% $\pm$ 0.033\%	& 0.178\% $\pm$ 0.057\% & 0.364\% $\pm$ 0.016\%	& 0.438\% $\pm$ 0.020\% \\
Brightness temperature ($K$)		  & 2065$^{+191}_{-155}$		& 2096$^{+206}_{-180}$	 & 2063 $\pm$ 36	                 & 2014 $\pm$ 41 \\	
Planet temperature ($K$)		  &  \multicolumn{4}{c}{2042 $\pm$ 50}  (joint solution) \\
Eclipse timing offset ($t-T_{\rm se}$)\tablenotemark{a} (min) 	&	fixed	to 0	&	fixed	to 0	& 	1.3 $\pm$ 1.4	& 	-1.2 $\pm$ 1.7		\\
Eclipse timing offset ($t-T_{\rm se}$)\tablenotemark{a} (min)	&		 \multicolumn{4}{c}{0.3 $\pm$ 1.3}  (joint solution)	
 \enddata
 \tablenotetext{a}{T$_{\rm se}$ is the predicted secondary eclipse time. }
\label{tab3}
\end{deluxetable*}

\begin{deluxetable}{lcc}
\tabletypesize{\scriptsize}
\tablecaption{Parameters for HAT-P-32A and HAT-P-32Ab}
\tablehead{
\colhead{Parameter} & \colhead{HAT-P-32A} & \colhead{Reference} 
} 
\startdata
~[Fe/H]			&	-0.04 $\pm$ 0.08	& 	1	\\
T$_{\rm eff}$ (K)		& 6269 $\pm$ 64		&	2 \\
R$_{*}$ (R$_{\odot}$)	& 1.219 $\pm$ 0.016	&	1\\
Distance (pc)		&	283 $\pm$ 5 &		1\\
\cutinhead{ HAT-P-32Ab}	
\sidehead{Transit Parameters}
R$_p$ (R$_{\rm Jup}$)		&	1.789 $\pm$ 0.025	&	1\\
a (AU)			& 0.0343 $\pm$ 0.0004	&	1\\
$i$ (deg)				& 88.9 $\pm$ 0.4	& 1	\\
T$_{\rm se, 3.6\micron}$ (BJD$_{\rm TDB}$)	&	2455855.57877 $\pm$ 0.00095	& 2\\
T$_{\rm se, 4.5\micron}$ (BJD$_{\rm TDB}$)	&	2455864.17701 $\pm$ 0.00118 & 2\\
 \sidehead{RV Model Parameters}
Period (days)			&	2.15000805 $^{+9.3e-07}_{-9.7e-07}	$	&	2\\
$T_0$ (BJD$_{\rm TDB}$)& 	2454420.44712 $^{+9.2e-05}_{-8.4e-05}$ 	&	2 \\
$e$					&	0.0072 $^{+0.0700}_{-0.0064}$			&	2 \\
$\omega$	(deg)			&	96 $^{+180.0}_{-11}$				& 	2 \\
$K$ (m s$^{-1}$)		&	110 $\pm$ 16						&	2 \\
$\gamma$ (m s$^{-1}$)	&	78 $^{+12}_{-13}$ 					& 	2 \\
$\dot{\gamma}$ (m s$^{-1}$ day$^{-1}$) & -0.048 $\pm$ 0.012			&	2 \\
jitter	(m s$^{-1}$)		&	67.2 $^{+9.6}_{-7.5}$				& 	2 \\
 \sidehead{RV Derived Parameters}
$e \cos \omega$		&	0.0004 $^{+0.0007}_{-0.0006}$		&	2 \\
$e \sin \omega$		&	0.0003 $^{+0.052}_{-0.010}$			& 	2
 \enddata
 \tablenotetext{1}{Hartman et al. 2011, ApJ, 742, 59}  
 \tablenotetext{2}{This work}   
\label{tab2}
\end{deluxetable}

\section{Discussion}
\label{discuss}

\subsection{Orbital eccentricity}
\label{ecc}

We can put a stringent constraint on the orbital eccentricity $e$ of the planet using the measured secondary eclipse timing from Section \ref{spitzer_lc}. 
The time delay due to light traveling across the orbit \citep[$\sim2a/c$, e.g.,][]{Kaplan2010} is $\sim$34.23 s for the hot Jupiter HAT-P-32Ab, so we expect to observe the secondary eclipse at an orbital  phase of 0.50018. Since the transit epoch reported in \citet{Hartman et al.2011}, a total of 667 and 671 orbital periods elapsed before our observations in the 3.6 and 4.5 $\mu$m bands, respectively. 
Taking the formal error of the transit epoch and the orbital period into account ($\sim$7.78 s and $\sim$0.0864 s, respectively), the cumulative errors resulting from the uncertain orbital period are 0.9605 min at 3.6 \micron~and 0.96624 min at 4.5 \micron. We add these uncertainties from the ephemerides in quadrature to our measurement uncertainty to obtain our final results for time delay from the predicted centers of eclipse for a circular orbit: 1.3 min $\pm$ 1.7 min at 3.6 \micron~and -1.2 min $\pm$ 2.0 min at 4.5 \micron. Therefore, the weighted-average timing delay from expected mid-occultation is $\Delta t \approx 0.3 \pm 1.3$ min (Table \ref{tab3}). 


For a complete constraint on the eccentricity, we incorporate the measured time delay in both bands into the RV orbital solution, using the RV data and procedures described in \citet{Knutson et al.2014},  as well as an additional 7 RV measurements obtained from Keck.
The resulting eccentricity of the orbit is  $e=$0.0072$^{+0.0700}_{-0.0064}$, which is consistent with a circular orbit at 1.1-$\sigma$.
Our secondary eclipse data constrain $|e \cos \omega|$ in these fits to be very close to zero, while $|e\sin\omega|$ is constrained primarily by the radial velocity measurements and spans a wider range of values.
We listed all the RV parameters  in Table \ref{tab2}.

A circular orbit for the planet is also preferred by  statistical tests 
 \citep[Lucy \& Sweeney test and BIC test, see][]{Hartman et al.2011},
 and  the short tidal circularization timescale of the system ($t_{\rm tidal} \sim3 -5$ Myr, much shorter than the $>$2 Gyr age of the system; \citet{Zhang et al.2013}). 
Due to  ambiguity in the radial velocity data,  \citet{Hartman et al.2011} provided two sets of solutions to the system, one with a fixed circular orbit $e=0$, and the other with a free floating $e$=0.163 $\pm$ 0.061. Our constraint on the eccentricity strongly prefers the $e=0$ solution. Therefore, the circular solution of the orbital and planetary parameters for HAT-P-32Ab \citep[middle column of Table 8 in][]{Hartman et al.2011} should be adopted as the formal parameters, which have already been used throughout this study. 
The  mass of HAT-P-32Ab is thus 0.860 $\pm$ 0.164 $M_{\rm Jup}$, and its radius is 1.789 $\pm$ 0.025 $R_{\rm Jup}$, making it the third largest transiting planet known to date.

\subsection{Atmospheric models for HAT-P-32Ab}

\begin{figure}[t]
\begin{center}
 \includegraphics[width=3in, angle=0]{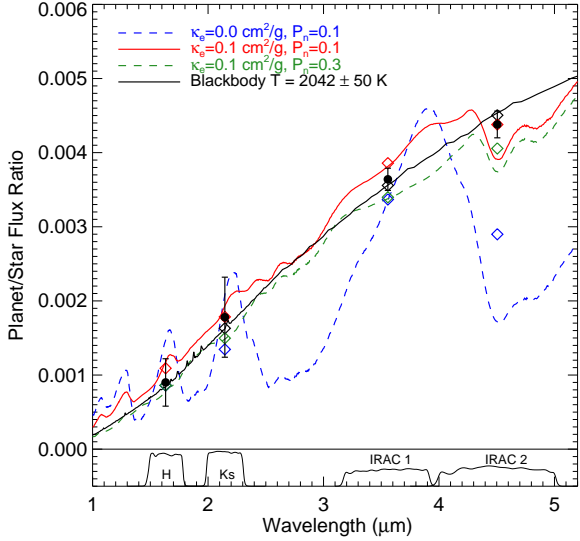}   
 \caption{Comparison of dilution corrected planet-to-star flux ratios with atmospheric models based on \citet{Burrows et al.2008}. The blue dashed line shows a nearly dayside-only redistribution model (close to 2$\pi$ redistribution) without upper-atmosphere heating (i.e., no temperature inversion). The green dashed line shows a model with a temperature inversion and moderate heat redistribution. The red line shows a model with a temperature inversion but very little redistribution. The black solid line indicates a blackbody model. Actual data points are shown as filled black dots with error bars. Colored diamonds indicate band-averaged models points.  
Normalized filter profiles of each bandpass are shown at the bottom. 
}
\label{burrows}
\end{center}
\end{figure}

\begin{figure}[t]
\begin{center}
 \includegraphics[width=3in, angle=0]{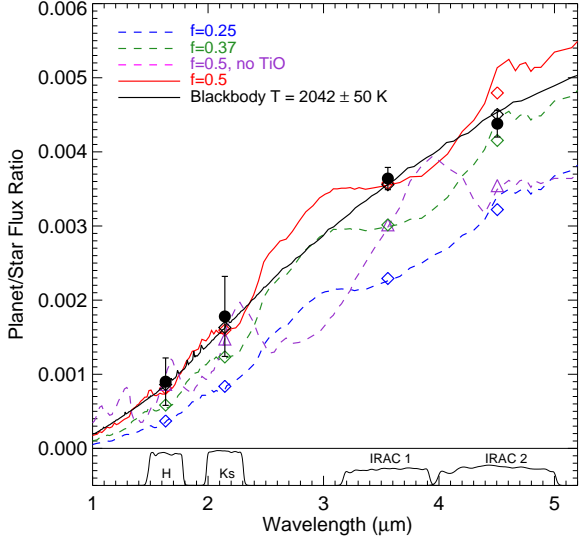}   
 \caption{Comparison of dilution corrected planet-to-star flux ratios with atmospheric models based on \citet{Fortney et al.2008}. The blue dashed line shows a planet-wide full redistribution model (4$\pi$ redistribution) with TiO in the upper atmosphere (i.e., has temperature inversion). The green dashed line shows a model with TiO (temperature inversion) and moderate heat redistribution. The solid red line shows a model with TiO but has only  day-side redistribution. The purple dashed line shows a day-side only model without TiO. The black solid line indicates a blackbody model. Actual data points are shown as filled black dots with error bars. Colored diamonds or triangles indicate band-averaged models points.  
Normalized filter profiles of each bandpass are shown at the bottom. 
  }
\label{fortney}
\end{center}
\end{figure}

We combine our final secondary eclipse depths in the $Spitzer$ 3.6 \& 4.5\micron~and WIRC $H$ \& $K_S$ bands  to compare with atmospheric models. We first fit a blackbody model to our broadband data and determine the effective temperature of the planet as well as its brightness temperatures in each bandpass (Table \ref{tab3}), using the PHOENIX  atmospheric models \citep{Husser et al.2013} for the host star with \teff=6200 K, $\log g$=4.5 cm s$^{-2}$ and [Fe/H]=0.0.
We find that our combined data are well-fit by a blackbody model with \teff=2042 $\pm 50$ K for the planet, which we show in Figure \ref{burrows} and \ref{fortney}.

We then compare our measurements with atmospheric models calculated as described in \citet{Burrows et al.2006, Burrows et al.2008} and \citet{Fortney et al.2008}, respectively.
Figure \ref{burrows} shows the dilution corrected planet-to-star flux ratios and comparison with  models based on  \citet{Burrows et al.2006, Burrows et al.2008}, assuming a plane-parallel atmosphere with local thermodynamic equilibrium (LTE),  solar abundance, and equilibrium chemistry. These models use a generalized absorption coefficient, $\kappa_e$, to represent the  unknown extra absorber in the upper atmosphere (stratosphere), which can cause extra heating and create a temperature inversion. The efficiency of energy redistribution is denoted by a dimensionless parameter $P_n$, with $P_n$ = 0.0 representing dayside-only redistribution (2$\pi$ redistribution), $P_n$ = 0.5 representing  a full redistribution over the planet (4$\pi$ redistribution) and other $P_n$ values representing intermediate level of redistribution. 
Figure \ref{burrows}  shows that the data prefer a model with extra upper-atmosphere heating ($\kappa_e = 0.1$ cm$^2$ g$^{-1}$), i.e., a temperature inversion, over a model without extra absorber, due to the high temperature and flux ratio in the 4.5 \micron~band. 
The ground-based data provide less leverage than the {\em Spitzer} data due to their larger uncertainties. The data also prefer a less efficient heat redistribution model with $P_n=0.1$ (red, total $\chi^2=2.47$) to a more efficient redistribution model of $P_n=0.3$ (green, total $\chi^2=6.12$), suggesting little recirculation to the planet's nightside.  
Nonetheless, we also note that a blackbody model provides a superior fit to the data ($\chi^2=0.9$), despite the fact that the 4.5 \micron~point is more consistent with the red model ($\kappa_e=0.1$ cm$^2$ g$^{-1}$, $P_n=0.1$).

Figure \ref{fortney} compares the dilution corrected planet-to-star flux ratios with models based on \citet{Fortney et al.2008}, also assuming a plane-parallel atmosphere with local thermodynamic equilibrium (LTE),  solar abundance, and equilibrium chemistry. Unlike the \citet{Burrows et al.2008} models, these models use gas-phase TiO as the upper atmospheric absorber and add it in chemical equilibrium to create a temperature inversion.  The efficiency of energy redistribution is denoted by a dimensionless parameter $f$, with $f$ = 0.25 representing dayside-only  (2$\pi$) redistribution, $f$ = 0.5 representing  a full-planet (4$\pi$) redistribution.
Figure \ref{fortney} shows that the model with TiO in the upper atmosphere and with very little heat redistribution ($f=0.5$) is more consistent with the data (red, total $\chi^2=5.68$). A model without TiO cannot match the data in the two $Spitzer$ bands (purple, total $\chi^2=39.2$), although it is consistent with the data in the $H$ \& $K_S$ bands. The blackbody model with \teff=2042 K still provides a better overall fit than any of the other models.
The 4.5 \micron~datum again provides the most leverage than other data points, while the other three points are consistent with both the 2$\pi$ TiO model and the blackbody model. 

Both sets of atmospheric models prefer an atmosphere with  a temperature inversion caused by high altitude absorber, and inefficient heat redistribution to the nightside of the planet. 
 The relatively high temperature of HAT-P-32Ab implies that it falls  near the inefficient end of the transition between efficient and inefficient redistribution on the trend found  by \citet{Cowan et al.2011} and \citet{Perez-Becker et al.2013}. Its low heat redistribution based on the models is therefore consistent with the observed correlation.
However, the planet host star HAT-P-32A has a moderately strong $\log R'_{HK}$  of -4.62 based on the analysis of \ion{Ca}{2} H and K line cores by \citet{Hartman et al.2011}. Thus it seems to be an exception to the hypothesis of \citet{Knutson et al.2010} that active stars with $\log R'_{HK} \gtrsim -4.9$  are likely to have non-inverted atmospheres as increased far UV flux from the star destroys the high altitude absorber in the upper atmosphere of the planet and suppresses temperature inversion. Nonetheless, we also note that the planet host star HAT-P-32A is at the boundary where the calibration for $\log R'_{HK}$ is highly uncertain (\teff $>6200$ K). \citet{Hartman et al.2011} also pointed out that HAT-P-32 does not show significant chromospheric emission in their \ion{Ca}{2} H \& K line cores. Therefore, this system may actually have weak far UV emission despite its large (but uncertain) $\log R'_{HK}$ value, and thus may not be an exception of the empirical correlation after all, as for the case of  XO-3 \citep{Knutson et al.2010}.

\section{Conclusions}
\label{summary}
We detected four secondary eclipses  of the hot Jupiter HAT-P-32Ab with the WIRC instrument at the Palomar Hale 200-inch telescope in the $H$ and $K_S$ bands, and with  {\em Spitzer} at 3.6 and 4.5 \micron. 
We characterized the flux-dependent nonlinearity of the HAWAII-2 detector of WIRC, and found that it can cause non-negligible effect to high precision photometry when the incident flux has large fluctuations in time, particularly at shorter wavelengths ($H$ or $J$ band) where the sky background is low. 
We also carried out AO imaging of the HAT-P-32AB system to resolve the planet host star from its nearby faint companion. We measured a separation of  2\farcs923 $\pm$ 0\farcs004 and a position angle of 110\fdg64 $\pm$ 0\fdg12 for the binary system.
By measuring the flux ratios of the stellar companion to the planet host star in $g'r'i'z'$ and NIR $H$ and $K_S$ bands, we determined a temperature of \teff=3565 $\pm$ 82 K for the companion, corresponding to an M1.5 dwarf. 
We extrapolated the flux ratios of the binary to the $Spitzer$ 3.6 and 4.5 \micron~bands based on their colors from the PHOENIX stellar atmosphere models. We then corrected the dilution to the secondary eclipse depths of the hot Jupiter in the four NIR bands to estimate the corresponding planet-to-star flux ratios. 
These corresponding flux ratios are 0.090 $\pm$ 0.033\%, 0.178 $\pm$ 0.057\%, 0.364 $\pm$ 0.016\%, and 0.438 $\pm$ 0.020\% in the $H$, $K_S$, and the $Spitzer$ 3.6 and 4.5 \micron~bands, respectively. 

By comparing the planet-to-star flux ratios with planetary atmospheric models, we found that both the \citet{Burrows et al.2008} models and the \citet{Fortney et al.2008} models prefer an atmosphere with temperature inversion caused by high altitude absorber and has inefficient heat redistribution to the nightside of the planet. 
The inefficient heat redistribution  is consistent with the trend found by previous studies of \citet{Cowan et al.2011} and \citet{Perez-Becker et al.2013}.
Given the moderately strong $\log R'_{HK}$ value of the planet host star, the hot Jupiter HAT-P-32Ab seems to be an exception to the correlation of \citet{Knutson et al.2010}. Nonetheless, because  the \teff~of the star is high, its   $\log R'_{HK}$  is largely uncertain and it may actually have low UV activities, which can make this system still consistent with the trend. 
Meanwhile, we also note that a blackbody model with $T_{p} =2042\pm 50$ K also fits the data well and cannot be distinguished from other models by our data. 

In addition, we measured a secondary eclipse timing offset of  0.3 $\pm$ 1.3 min from the predicted mid-eclipse time. We combined this with new RV data from Keck and those from \citet{Knutson et al.2014} to put more stringent constraint on the eccentricity of the hot Jupiter's orbit. 
We found  $e$ = 0.0072 $^{+0.0700}_{-0.0064}$, which is  consistent with a circular orbit at the 1.1-$\sigma$ level.  The presence of both a radial velocity acceleration and a separate, directly imaged companion in this system might point to a complex dynamical history that could have resulted in the planet's extreme spin-orbit misalignment at present.
A circular orbital solution  also makes HAT-P-32Ab the third largest planet known to date.  
Because high-eccentricity dynamical migration mechanisms such as the Kozai migration can be very slow ($\sim$Gyrs) \citep{Wu et al.2003}, this planet might have reached its present configuration relatively recently, which may partially explain its abnormally large radius. 

In addition to the M1.5 companion, \citet{Knutson et al.2014} detected a radial velocity trend of $-33\pm10$m s$^{-1}$yr$^{-1}$ in the system. The trend, however, cannot be explained by the  companion star due to its large separation of $\sim$830 AU, but requires an inner body at 3.5-21 AU from the planet host star with a projected mass ($M\sin i$) between 5-500 $M_{Jup}$. At a distance of 283 pc, the inner body has a angular separation between 0\farcs012 - 0\farcs074 and is unresolved by our AO imaging. 
{Therefore, if the unseen inner body is a stellar companion close to $\sim$500 $M_{Jup}$ (i.e., a late K dwarf), it would further dilute the transit and secondary eclipse depths of HAT-P-32Ab, making its radius even larger.}
Nonetheless, this potential unseen body would not significantly affect the characterization of the hot Jupiter's atmosphere due to its small and negligible flux contribution. 

Due to the large apertures used in our photometry (and correspondingly much more background noise in the photometry), and the correlated instrumental systematics cause by telescope astigmatism, our ground-based photometry of HAT-P-32Ab in this study was not able to provide strong constraints on the atmospheric models. 
{Future observations  with stabilized PSFs \cite[e.g., using a diffuser,][]{Zhao et al.2014} and a newer generation of IR detectors (e.g., the HAWAII-2RG series) will be able to mitigate the correlated systematics at the instrument level and achieve precisions of $\lesssim$100 parts-per-million for better characterization of planetary atmospheres.  
Although the WIRC detector used in this study is now defunct, it may be upgraded to a new science-grade HAWAII-2RG array and  allow better performance in the future.}

\acknowledgments
We thank the anonymous referee for valuable comments for the paper. 
We thank the Palomar  staff for their help with the observations. 
M.Z. is supported by funding from NASA Origins of Solar Systems grant NNX14AD22G and the Center for Exoplanets
 and Habitable Worlds at the Pennsylvania State University. 
  The Center for Exoplanets and Habitable Worlds is supported by the
 Pennsylvania State University, the Eberly College of Science, and the
 Pennsylvania Space Grant Consortium.
 
J.G.O. receives support from the National Science Foundation's Graduate Research Fellowship Program.
H.N. acknowledges funding support from the Natural Science and Engineering Research Council of Canada. 
A.B. acknowledges support in part under NASA HST grants HST-GO-12181.04-A, HST-GO-12314.03-A, HST-GO-12473.06-A, and HST-GO-12550.02, and JPL/Spitzer Agreements 1417122, 1348668, 1371432, 1377197, and 1439064.
S.H. acknowledges support from the NASA Sagan Fellowship at California Institute of Technology.  
C.B. acknowledges support from the Alfred P. Sloan Foundation. 

The Palomar Hale 200 inch Telescope is operated by Caltech and the Jet Propulsion Laboratory.

The Robo-AO system is supported by collaborating partner institutions, the California Institute of Technology and the Inter-University Centre for Astronomy and Astrophysics, by the National Science Foundation under Grant Nos. AST-0906060, AST-0960343, and AST-1207891, by a grant from the Mt. Cuba Astronomical Foundation and by a gift from Samuel Oschin. 

Some of the data presented herein were obtained at the W.M. Keck Observatory, which is operated as a scientific partnership among the California Institute of Technology, the University of California and the National Aeronautics and Space Administration. The Observatory was made possible by the generous financial support of the W.M. Keck Foundation.

This research has made use of the Exoplanet Orbit Database
and the Exoplanet Data Explorer at exoplanets.org. 

{\it Facilities}: \facility{Hale (WIRC), Spitzer (IRAC), PO:1.5m (Robo-AO), Keck:II (NIRC2)}.

\appendix

\section{Characterization of the flux-dependent nonlinearity of the HAWAII-2 detector}
\label{appendix}

Unlike classical CCD-type detectors which have nonlinearity as a function of  photons counts hitting the detector (i.e., count-dependent nonlinearity or gain nonlinearity), IR HgCdTe detectors (such as the NICMOS, HAWAII, and HAWAII-RG detectors)  have another type of nonlinearity that is not only  dependent on the total photon counts, but also dependent on the count-rate or incident flux level. That is, a fainter source hitting the detector with a lower photon rate gives a different nonlinearity response from a brighter source with a higher photon rate even if they give the same total number of counts. This count-rate dependent nonlinearity, also known as flux-dependent nonlinearity, or ``reciprocity failure" in photography, is relatively well known among engineers working on infrared detectors, but is less well-known among observational astronomers. It was first discovered in $HST$ NICMOS data \citep{Bohlin et al.2005} and characterized by \citet{Bohlin et al.2006} and \citet{deJong et al.2006}, for which a nonlinearity of $\sim$6\% per decade of flux change  was reported for NICMOS. Later studies also investigated this effect on newer generation HgCdTe IR detectors from Teledyne Imaging Sensors, such as the $HST$ WFC3 IR detector and the HAWAII-RG family detectors \citep{Hill et al.2010, Deustua et al.2010, Biesiadzinski et al.2011a}. Despite the still-unclear physical mechanism behind the flux-dependent nonlinearity,  studies have found that this effect is common in HgCdTe detectors and varies from detector to detector \citep{Biesiadzinski et al.2011a}. Nonetheless, the newer generation HAWAII-RG series have significantly lower flux-dependent nonlinearity than older generation detectors, and this effect is generally not wavelength-dependent \citep{Biesiadzinski et al.2011a}.

Characterizing and calibrating the flux-dependent nonlinearity for HgCdTe IR detectors is important for high precision photometric measurements such as measuring transmission spectra or secondary eclipses of exoplanets, as the signatures of interest are often extremely small and dominated by noise or systematics. 
In principle, because these measurements only target  differential flux changes as a function of time, the differential nonlinearity effects are minimal and usually negligible as long as the incident flux and PSFs are stable, and the PSFs are well sampled by many pixels (so minor variations can be averaged out). 
However, because ground-based observations usually suffer from large flux variations from pixel to pixel due to atmospheric variations or time-varying instrumental effects (e.g., varying bright spots on the PSF due to telescope astigmatism), it is necessary to understand the impact of this effect. 

We characterized the flux-dependent nonlinearity of the HAWAII-2 detector of Palomar Hale/WIRC (an earlier generation of the HAWAII-2RG family) using dome flats with the detector installed on the telescope. 
To ensure the stability of the flat lamp, we kept the power of the lamp stable and unchanged, and simulated different incident flux levels by changing the amount of mirror cover opening.
At each fixed incident flux level, we took many sets of exposures with increasing exposure times to sample the nonlinearity curve.
We took minimum exposures interspersed between each set of measurements to measure and track the incident flux level of each nonlinearity curve. This ensures that small flux variations of the flat lamp are properly corrected. 
 Although the absolute flux level cannot be measured accurately in this manner due to existing nonlinear counts from the detector, it still allows us to track the relative differential nonlinearity changes at each observed flux level. 
By sampling the nonlinearity response of the whole detector at different incident flux levels and comparing with the actual counts, we can construct a nonlinearity response map as a function of  observed flux  and total  counts per pixel  -- this allows us to correct for the nonlinearity of each pixel based on its measured counts and average flux. The measured-to-actual counts ratio, i.e., the nonlinearity response, can be denoted as:
\begin{equation}
\eta_{i,n} = f(\bar F_{i,n} ~,~ c_{i,n} ),
\end{equation}
 where $\bar F_{i, n}$ is the mean flux of pixel $i$ in image $n$, and $c_{i,n}$ is the total measured counts of pixel $i$ in image $n$. 
 The mean flux of pixel $i$ in image $n$ can be approximated using its total counts and the exposure time  of the image as long as the exposure time is not too long (e.g., $\sim10$ s):
 \begin{equation}
 \bar F_{i,n} = \frac{c_{i,n}}{t_{i,n}}.
 \end{equation}

Since all pixels have the same exposure time in each image, i.e., $t_{i,n} = t_{n}$,  we can take this advantage and convert the response from a function of flux and counts into a function of exposure time and counts. Therefore, for fixed exposure time $t_{i,n} = t_0$, as for the case of our secondary eclipse observations, the nonlinearity function can be further simplified to:
\begin{equation}
\eta_{i,n} = f(\frac{c_{i,n}}{t_0}, c_{i,n} ),
\end{equation}
which is essentially a function of  pixel counts $c_{i,n}$ when the exposure time is fixed. Note that because of the difficulty in measuring the actual flux level of each individual pixel  in realtime when using the same IR detector that needs to be calibrated, this method only provides an approximation to  the flux-dependent nonlinearity effect.

\begin{figure}[t]
\begin{center}
 \includegraphics[width=3in, angle=0]{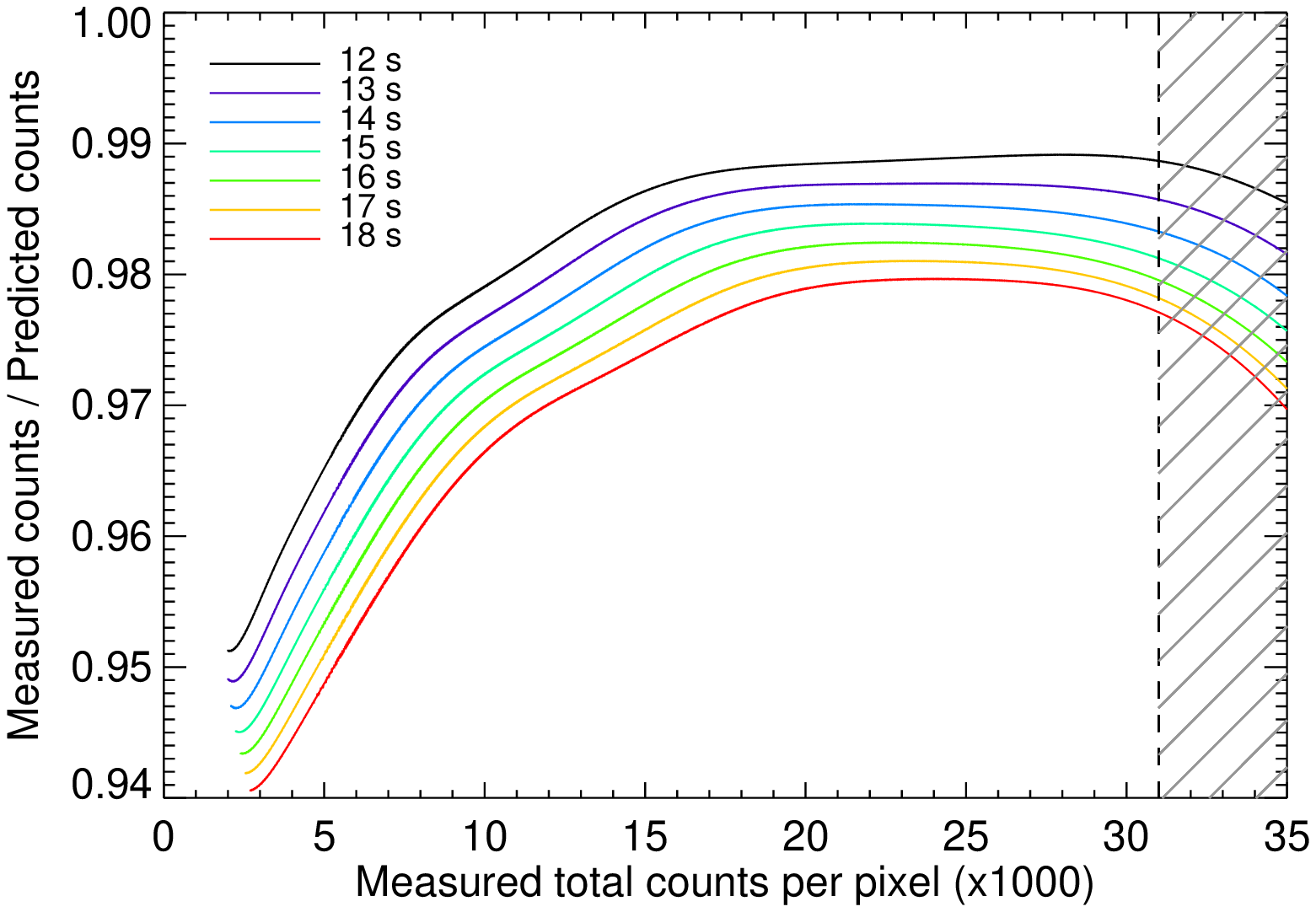}   
\hspace{0.35in}
 \includegraphics[width=3in, angle=0]{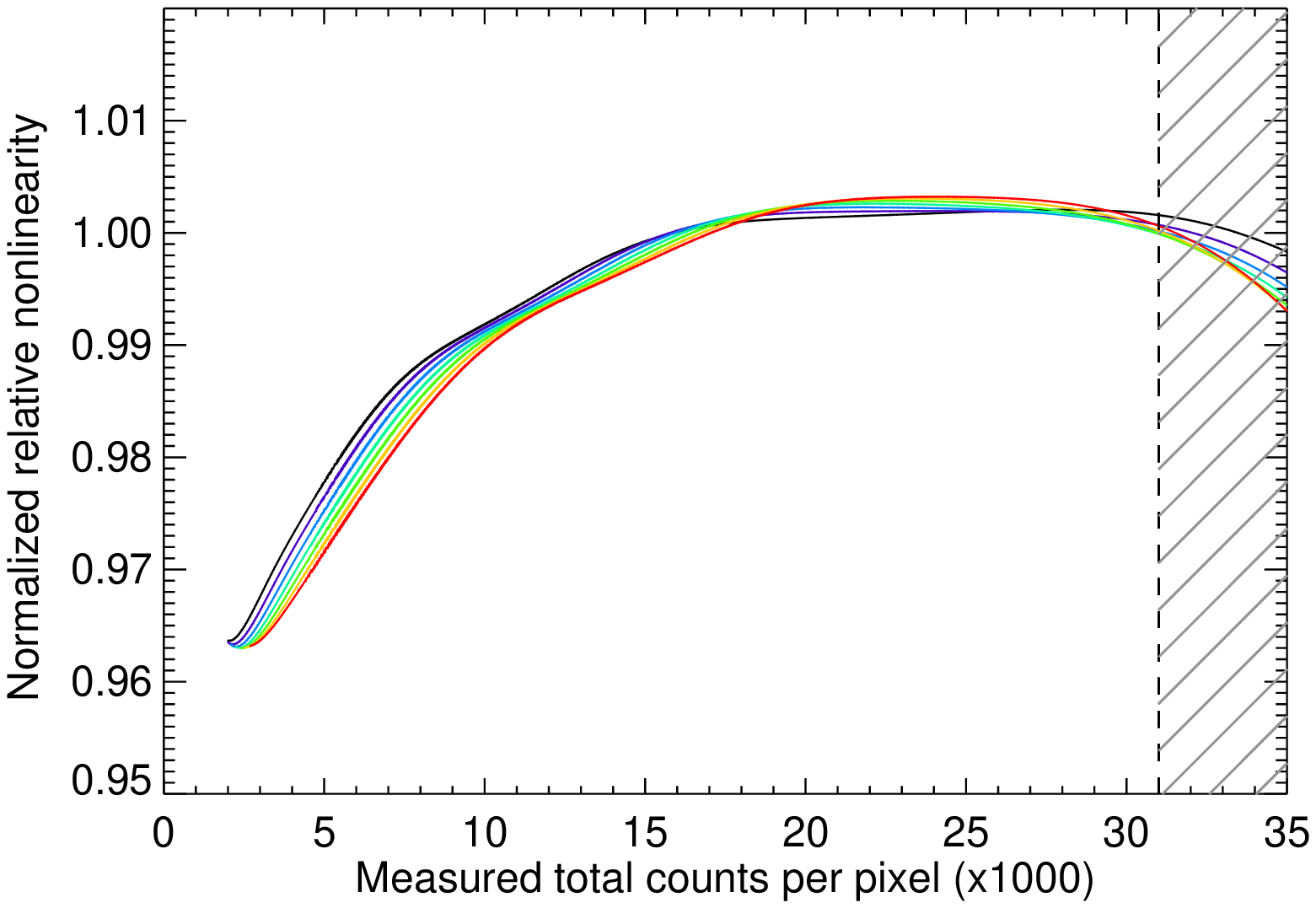}   
 \caption{Left: Nonlinearity of  WIRC's HAWAII-2 detector measured in the form of Counts$_{\text{measured}}$/Counts$_{\text{predicted actual}}$ as a function of pixel counts  and exposure times (denoted as colored lines). Corresponding exposure times for the colored lines are labeled in the top left. The shaded area indicates the region where the detector begins to saturate and  measurements become unreliable. Note the average flux level changes along each colored curve. At any given total counts, shorter exposures correspond to higher flux levels and deviate less from the actual counts.
Right: We normalize the nonlinearity curves in the left panel and average them to create a global, relative nonlinearity curve since the shapes of the curves are similar. However, this averaged curve only applies to images taken with the same exposure time. Images taken with difference exposure times cannot be analyzed together due to absolute differences in their nonlinearity responses. 
  }
\label{nonlinear}
\end{center}
\end{figure}
 
 Figure \ref{nonlinear} shows our measured flux-dependent  nonlinearity of the HAWAII-2 detector of WIRC as a function of pixel counts and exposure time (shown as different colors).
Due to limited dynamic range of the incident flux (limited by the brightness of the lamp and the minimum controllable opening angle of the mirror cover), we could only measure the nonlinearity curves for exposure times between 12 s and 18 s reliably. 
 The nonlinearity  curves are  clearly flux-dependent in the figure. Given the same number of total counts, high incident flux levels (short exposures) cause less deviation from unity than that of low flux levels (long exposures), a trend that is consistent with previous studies \citep{deJong et al.2006, Biesiadzinski et al.2011a}. The shape of each nonlinearity curve closely resembles each other despite their difference in absolute  nonlinearity levels. All curves have a ``ramp" at low counts ($\lesssim$15 K) and a ``plateau" between $\sim15$ K counts and 30 K counts, and start to drop off at $>$30 K when the counts are close to saturation. This suggests that for images with high background levels (BG), such as the $K_S$ band images (BG$>$12 K) in our observations, most of the pixels are in the ``plateau" region, and thus have negligible differential nonlinearity effects for differential photometry. In contrast, images in the $H$ band have more significant nonlinearity  differences between  the wings and peaks of each stellar PSF due to their low background levels and  large differences in counts. 
 
 Since we only measure differential photometry in our secondary eclipse observations, we can take the advantage of the similar  nonlinearity shapes and average them to create a normalized differential nonlinearity curve that is universal to all exposure times. The right panel of Figure \ref{nonlinear} overplots the normalized nonlinearity curves together and shows that they  indeed have very consistent shapes. We therefore fit a polynomial function to the averaged curve. 
The standard deviation of all the curves at each data point is taken as the measurement error in the fit.   We again use BIC to select the best model. An 8th order polynomial gives the smallest BIC ($\Delta BIC>20$) and is strongly preferred. 
 The resulting global nonlinearity approximation is:
 \begin{equation}
 \eta =  0.9741 -0.1522x +0.6556x^2  -1.0664x^3 + 0.9425x^4 -0.4878x^5 + 0.1472x^6 -0.02398x^7 + 0.00163x^8 , 
  \end{equation} 
assuming all pixels on the detector have the same response, where $x = \rm{counts}/10^4$ is the scaled counts of  each pixel. 
We applied this nonlinearity curve to our WIRC data in the $H$ band based on the assumption that the detector response remains similar for exposure times lower than 12s.
We then divided each pixel's measured counts by its corresponding nonlinearity approximation in Section \ref{wirc}.
We also applied this nonlinearity approximation to the $K_S$ band data. However, the resulting difference in $K_S$ is negligible, as expected from its high background levels. 
This global nonlinearity curve is applicable to all WIRC data obtained using the same HAWAII-2 detector and the same exposure times. However, since the flux-dependent nonlinearity varies from detector to detector, this function form cannot be applied to other instruments. Nonetheless, the measurement method described here can be applied to similar  characterizations of HgCdTe IR detectors. Meanwhile, we also want to point out that tests conducted with adjustable, highly stable lamps with well known incident flux levels will provide the best results, and thus are highly recommended.\\ 

  
 \vspace{0.2in}
 






\end{document}